\documentclass[11pt]{article}
\usepackage{amsfonts,amstext,amsmath,amssymb,setspace,amsthm,mathtools,caption2,color,enumerate, physics, mathtools, diagbox, bm, adjustbox}
\usepackage{graphicx, subfigure}
\usepackage[shortlabels]{enumitem}
\usepackage{lscape, bbm} 
\usepackage{calrsfs}
\usepackage[autostyle]{csquotes}
\usepackage{mathrsfs} 
\usepackage[toc,page]{appendix}
\usepackage[square,numbers]{natbib}
\usepackage[colorlinks]{hyperref}
\hypersetup{citecolor= blue}
\hypersetup{linkcolor= red}
\hypersetup{urlcolor= blue}
\usepackage{algorithm}
\usepackage{algorithmic}
\usepackage[hang, flushmargin, symbol]{footmisc}
\allowdisplaybreaks
\theoremstyle{plain} 
\newtheorem{theorem}{Theorem}
\newtheorem{lemma}{Lemma}
\newtheorem{corollary}{Corollary}
\theoremstyle{definition} 
\newtheorem{definition}{Definition}
\newtheorem{remark}{Remark}

\newcommand{\bi}{\begin{itemize}} 
	\newcommand{\ei}{\end{itemize}}
\newcommand{\be}{\begin{enumerate}} 
	\newcommand{\ee}{\end{enumerate}}
\def\ba#1\ea{\begin{align}#1\end{align}} 
\def\bdef#1\edef{\begin{definition}#1\end{definition}} 
\def\bthm#1\ethm{\begin{theorem}#1\end{theorem}} 
\def\blem#1\elem{\begin{lemma}#1\end{lemma}} 
\def\bcor#1\ecor{\begin{corollary}#1\end{corollary}} 
\def\bpf#1\epf{\begin{proof}#1\end{proof}} 
\def\brem#1\erem{\begin{remark}#1\end{remark}} 
\def\bex#1\eex{\begin{example}#1\end{example}} 
\providecommand{\abs}[1]{\lvert#1\rvert} 
\renewcommand{\b}{\textbf} 
\def\({\left(}  
\def\){\right)}

\def\Y{\mathcal{Y}}

\begin{document}
	\title{\textbf{An EM algorithm for absolutely continuous Marshall-Olkin bivariate Pareto distribution with location and scale}}
	\date{}

\maketitle
%
\begin{center}
\null\vskip-2.0cm
\author{  \textbf{ Biplab\ Paul}$^{1}$, \textbf{Arabin\ Kumar\ Dey}$^{2}$
	\\
	$^{1}$Department of Statistics, University of Haifa, Israel\\
	$^{2}$ Department of Mathematics, IIT Guwahati, India
}
\end{center}
\begin{center}
\today
\end{center}
\footnotetext{Correspondence author's Email address: \url{paul.biplab497@gmail.com} (Biplab Paul)} 
%

\begin{abstract}
In this paper, we have considered a Block-Basu type bivariate Pareto distribution. Here in the standard manner, first Marshall-Olkin type singular bivariate distribution has been constructed, and then by taking away the singular component similar to the Block and Basu model, an absolute continuous BB-BVPA model has been constructed. Further, the location and scale parameters also have been introduced. Therefore, the model has seven parameters. Different properties of this absolutely continuous distribution are derived. Since the maximum likelihood
estimators of the parameters cannot be expressed in a closed form, we propose to use an EM algorithm to compute the estimators of the model parameters. Some simulation experiments have been performed for illustrative purposes. The model is fitted to rainfall data in the context of landslide risk estimation.
\end{abstract}

\section{Introduction}
\label{s1}
%
Statistical inference for extreme values has received extensive attention over the past couple of decades. They are extensively discussed by  \citet{coles2001introduction}, \citet{embrechts2013modelling}, \citet{kotz2000extreme}, \citet{reiss2005statistical}, and so on. Pareto distribution plays an important role in extreme value (EV) theory. It is noted that there is an intimate relation between the Pareto type $II$ distribution and the Pickands generalized Pareto model \cite[see, e.g.,][]{arnold2015}. Different variants of the classical Pareto distributions are typically used to analyze income and wealth data. Bivariate Pareto has a wide application in many applied areas, including finance,  failure times, income and wealth modeling, insurance, environmental sciences, internet network, modeling of birth rates and infant mortality rates, reliability, etc. In a financial setting, multivariate Pareto distribution can be used to model the dependent risks associated with lines of business \cite[see, e.g.,][]{vernic2011tail}. This distribution might be used to estimate system reliability in stress-strength setting \cite[see, e.g.,][]{hanagal1996multivariate}. Multivariate Pareto distribution is also useful to measure the impact of extreme events, which often depend on not just a single component but the combined behavior of several components of interest.

Even though different types of univariate Pareto distributions \cite[see, e.g.,][]{arnold1989bayesian, arnold1998bayesian} have been analyzed extensively, there is a lack of methods for analyzing multivariate Pareto distributions. In 1962, \citet{mardia1962multivariate} first systematically studied multivariate Pareto distribution where marginals are Pareto type $I$ distribution with common shape parameter. \citet{arnold2015} proposed multivariate Pareto distribution type $II$ using same procedure as in \citet{mardia1962multivariate}. The Bivariate Lomax distribution proposed by \citet{lindley1986multivariate} is another popular bivariate Pareto distribution, which also has Lomax marginals. A detailed description of multivariate Pareto distribution can be found in \cite{arnold2015}. In this context, a book by \citet{kotz2004continuous} can be a good reference. However, inference methods for various forms of multivariate Pareto distributions have been somewhat restricted. The main reason for this limitation is a lack of appropriate models which predict multivariate Paretian behavior. Recently \citet{asimit2016statistical} proposed Marshall-Olkin bivariate Pareto (MOBVPA) distribution using the same procedure as in \cite{marshall1967multivariate} and used the Expectation-maximization (EM) algorithm to estimate the unknown parameters. \citet{DeyPaul:2019} also described other innovative variations of the EM algorithm to estimate the unknown parameters of the MOBVPA model. MOBVPA distribution works quite effectively to analyze data when some of the two components of standardized dataset (location-scale transformation) take equal values because it is a singular distribution. In real-life datasets, we do not know the exact value of location and scale parameters, so it is quite impossible to know whether the standardized dataset has equal components. Therefore, this MOBVPA model with location and scale leads to misleading results when the standardized dataset does not have equal components. 

In 1974, \citet{block1974continuous} proposed the bivariate exponential distribution (BBBE) from the Marshall-Olkin bivariate exponential (MOBE) distribution by removing the singular part and retaining only the absolutely continuous part. This BBBE distribution is one of the most popular and widely used absolutely continuous bivariate distributions because, unlike MOBE, it enjoys all the properties of an absolutely continuous distribution. Interestingly, marginals are not the exponential distributions. 

Along the same line as BBBE distribution \cite[see, e.g.,][]{block1974continuous}, Block-Basu bivariate Pareto (hereafter called BB-BVPA) distribution has been defined. This distribution has been obtained from the MOBVPA distribution by removing the singular part, which makes the BB-BVPA distribution an absolutely continuous bivariate distribution. Similar to the MOBVPA model, BB-BVPA also has seven parameters. Although extensive works \cite[see, e.g.,][]{kundu2008generalized, kundu2009bivariate,  kundu2010class, kundu2011absolute, kundu2015absolute, mirhosseini2015new} have been done on the absolutely continuous version of MOBE and MOBW models, not that much of attention has been paid on BB-BVPA model. The reason might be, computationally it may not be very tractable, especially in the presence of both location and scale parameters. In fact, computing the maximum likelihood estimators (MLEs) of the unknown parameters of the BB-BVPA model in the location and scale parameters is not a trivial issue. Only recently, \citet{PaulDeyKundu:2018} proposed a Bayesian inference procedure for BB-BVPA distribution without location and scale parameters. We introduce location and scale parameters to the BB-BVPA model; hence it brings more flexibility. Unlike the bivariate Pareto distribution by \citet{arnold2015}, \citet{lindley1986multivariate} and \citet{mardia1962multivariate}, the marginals of BB-BVPA do not have a common shape parameter due to the presence of multiple shape parameters in the BB-BVPA model, which gives more flexibility. Although the marginals of BB-BVPA are not Pareto type $II$ in general, the shape of the PDF of marginals is very similar to the PDF of a Pareto type $II$ distribution. Interestingly, for non-negative location parameters, the distribution of these marginals are weighted distributions corresponding to Pareto type $II$ distribution, and, hence, the properties of weighted distribution \cite[see, e.g.,][]{nanda1999some} are also available here. Moreover, without any restriction on model parameters, we have shown that the BB-BVPA model without location and scale parameters has \textit{total positivity of order $2$} ($TP_2$). This distribution also has some properties which have proved useful for reliability applications. To the best of our knowledge, the formulation and estimation methodologies of the BB-BVPA distribution with location and scale parameters are yet to be developed. 

In this article, different properties and different computational
issues associated in computing the unknown parameters of the BB-BVPA distribution are discussed. First, we consider the computation of MLEs of the seven unknown parameters of the BB-BVPA model. As expected, the MLEs cannot be obtained in explicit form. To compute the MLEs directly, one needs to solve a multi-dimensional optimization problem. It is observed that the EM algorithm can be used quite effectively to compute MLEs of parameters of BB-BVPA. At each EM step (iteration), one needs to solve only a one-dimensional optimization problem, and we have proposed a simple procedure to solve this problem. It should be mentioned that the usual gradient descent does not work as the bivariate likelihood is a discontinuous function for the location and scale parameters. We suggest a novel way to handle all related computational problems which work most efficiently. The first contribution of this article is to implement the EM algorithm for the three-parameter BB-BVPA model without location and scale parameters, where a crucial modification of the EM algorithm is suggested to make the algorithm work for any range of parameters. The other contribution is an exploration of an efficient EM algorithm that will work in the case of the seven-parameter setup. The suggested structure combines several ideas, including previously suggested modification of the three-parameter setup. The distribution can be used quite effectively for the data transformed via the peak over threshold method, especially when the transformed dataset does not have equal component observations. The dependence structure of this absolute continuous version can also be described by the well-known Marshall-Olkin copula \cite[see, e.g.,][]{marshall1996copulas, nelsen2007introduction}. It should be noted that the methodology proposed here works quite well for moderately large samples. We also propose to use a naive construction of confidence interval for the parameters. 

We illustrate the usefulness of this BB-BVPA model by fitting it to extreme precipitation data and by showing how the results could be used to estimate risks for landslides. Extreme precipitation has a number of potentially hazardous consequences, such as flooding, plugged drainage systems, wasted crops and landslides. This analysis also helps to plot insurance policies to compensate for the losses or damage caused by floods, landslides, etc.

The organization of the paper is as follows. A brief description of the formulation and different properties of BB-BVPA distribution are given in Section~\ref{s2}. Section~\ref{s3} is kept for different extensions of the EM algorithm of BB-BVPA distribution. We discuss the construction of confidence interval in Section~\ref{s5}. Some simulated datasets are analyzed in Section~\ref{s6} for illustrative purposes. Section~\ref{s7}, we fit our BB-BVPA model to rainfall data in the context of landslide risk estimation. Finally, Section~\ref{s9} contains a brief summary of the work and some concluding remarks.

\section{Formulation of Block-Basu bivariate Pareto distribution}
\label{s2}
If a random variable $X$ has a univariate Pareto type $II$ distribution
with the location, scale and shape parameters as $\mu\in \mathbb{R}$, $\sigma>0$ and $\alpha>0$, respectively, then for $x > \mu$, the probability density function (PDF) and survival function (SE) are defined as follows;

\begin{equation}
	\label{2e1}
	f_{PA}(x ; \mu, \sigma, \alpha) = \dfrac{\alpha}{\sigma}\Big(1 + \dfrac{x - \mu}{\sigma}\Big)^{-\alpha - 1},~~ S_{PA}(x ; \mu, \sigma, \alpha) = \Big(1 + \frac{x - \mu}{\sigma}\Big)^{-\alpha}, 
\end{equation}
respectively. From now on a Pareto type $II$ distribution with the PDF \eqref{2e1} will be denoted by $PA(II)(\mu, \sigma, \alpha)$. Let us consider $U_0$ follows $(\sim ) PA(II)(0, 1, \alpha_0)$, $U_1 \sim PA(II)(\mu_1, \sigma_1, \alpha_1)$ and $U_2 \sim PA(II)(\mu_2, \sigma_2, \alpha_2)$ and also they are mutually independent. Define $X_{1} = \min\{ \sigma_{1} U_{0} + \mu_{1}, U_{1} \}$ and $X_{2} = \min\{ \sigma_{2} U_{0} + \mu_{2}, U_{2} \}$, then the bivariate random variable $(X_1,X_2)$ has MOBVPA distribution with parameters $(\mu_{1}, \mu_{2}, \sigma_{1}, \sigma_{2}, \alpha_{0}, \alpha_{1}, \alpha_{2})$ and it will be denoted from now on as $MOBVPA(\mu_{1}, \mu_{2}, \sigma_{1},$ $ \sigma_{2}, \alpha_{0}, \alpha_{1}, \alpha_{2})$. The joint PDF of $(X_1, X_2)$ is
\begin{equation}
	\label{2e2}
	f(x_1,x_2)=
	\begin{cases}
		f_1(x_1,x_2),~\text{if $\frac{x_1-\mu_1}{\sigma_1}$ \textless $\frac{x_2-\mu_2}{\sigma_2}$}
		\\
		f_2(x_1,x_2),~ \text{if $\frac{x_1-\mu_1}{\sigma_1}$ \textgreater $\frac{x_2-\mu_2}{\sigma_2}$}\\
		f_{0}(x), ~ \text{if $\frac{x_1-\mu_1}{\sigma_1}$ = $\frac{x_2-\mu_2}{\sigma_2}$}=x,
	\end{cases} 
\end{equation}
where 
\begin{equation*}
	\begin{split}
		f_1(x_1,x_2)&=\frac{\alpha_1 (\alpha_0+\alpha_2)}{\sigma_1 \sigma_2}\Big(1+\frac{x_2-\mu_2}{\sigma_2}\Big)^{-\alpha_0-\alpha_2-1}\Big(1+\frac{x_1-\mu_1}{\sigma_1}\Big)^{-\alpha_1-1}\\
		f_2(x_1,x_2)&=\frac{\alpha_2 (\alpha_0+\alpha_1)}{\sigma_1 \sigma_2}\Big(1+\frac{x_2-\mu_2}{\sigma_2}\Big)^{-\alpha_2-1}\Big(1+\frac{x_1-\mu_1}{\sigma_1}\Big)^{-\alpha_0-\alpha_1-1} \\
		f_0(x)&= \alpha_0(1+x)^{-\alpha_0-\alpha_1-\alpha_2-1}.
	\end{split}
\end{equation*}
When $\mu_1 = \mu_2 = 0$ and $\sigma_1 = \sigma_2 = 1$, we call it three-parameter MOBVPA distribution and  will be denoted by $MOBVPA(\alpha_0,\alpha_1,\alpha_2)$. Note that Block-Basu bivariate Pareto distribution can be obtained from MOBVPA distribution by removing the singular part and keeping only the continuous part. The joint PDF of BB-BVPA distribution is
\begin{eqnarray}
	\label{2e3}
	f_{BB}(y_{1}, y_{2}) & =  
	\begin{cases}
		c f_{1}(y_{1}, y_{2})~~\text{if $\frac{y_{1} - \mu_1}{\sigma_1} < \frac{y_{2} - \mu_2}{\sigma_2}$}\\ c f_{2}(y_{1}, y_{2})~~\text{if $\frac{y_{1} - \mu_1}{\sigma_1} > \frac{y_{2} - \mu_2}{\sigma_2}$},
	\end{cases} 
\end{eqnarray}
where c is a normalizing constant and $c = \frac{\alpha_{0} + \alpha_{1} + \alpha_{2}}{\alpha_{1} + \alpha_{2}}$. Therefore, the joint PDF of bivariate random variable $(Y_1, Y_2) $ can be written as \eqref{2e3} and it will be denoted as $BBBVPA(\mu_{1}, \mu_{2}, \sigma_{1}, \sigma_{2}, \alpha_{0}, \alpha_{1}, \alpha_{2})$. The joint PDF of $(Y_1, Y_2)$ is unimodal. Surface and contour plots of $f_{BB}(y_{1}, y_{2})$ for different values of parameters are shown in Figure~\ref{fig1_ch3}. When $\mu_1 = \mu_2 = 0$ and $\sigma_1 = \sigma_2 = 1$, we call it three-parameter BB-BVPA distribution and denote this distribution as $BBBVPA(\alpha_0,\alpha_1,\alpha_2)$.
\subsection{Properties} Different properties of this BB-BVPA distribution are studied here. First, we provide the marginal and conditional distributions of the BB-BVPA distribution.

\begin{theorem}
	\label{th1}
	If $(Y_1, Y_2)$ $\sim$ $BBBVPA(\mu_1, \mu_2, \sigma_1, \sigma_2, \alpha_0, \alpha_1, \alpha_2)$, then the marginal PDFs of $Y_1$ and $Y_2$ are as follows
	\begin{eqnarray*}
		f_{Y_{1}}(y_{1}) = cf_{PA}(y_{1}; \mu_{1}, \sigma_{1}, \alpha_{0} + \alpha_{1}) - \frac{\alpha_{0}}{(\alpha_{1} + \alpha_{2})}f_{PA}(y_{1}; \mu_{1}, \sigma_{1}, \alpha_{0} + \alpha_{1} + \alpha_{2})
		\label{proper1}
	\end{eqnarray*}
and
	\begin{eqnarray*}
		f_{Y_{2}}(y_{2}) =  cf_{PA}(y_{2};\mu_{2}, \sigma_{2}, \alpha_{0} + \alpha_{2}) - \frac{\alpha_{0}}{(\alpha_{1} + \alpha_{2})}f_{PA}(y_{2}; \mu_{2}, \sigma_{2}, \alpha_{0} + \alpha_{1} + \alpha_{2}),
		\label{proper2}
	\end{eqnarray*}
	respectively, where $c = \frac{\alpha_{0} + \alpha_{1} + \alpha_{2}}{\alpha_{1} + \alpha_{2}}$.
\end{theorem}
\begin{proof}
	They can be obtained by the definition of marginal distribution from the joint distribution.
\end{proof}

\begin{figure}[h]
	\centering
	\subfigure[Probability density functions.]{\label{pdf}
		\includegraphics[height=70mm, width=70mm]{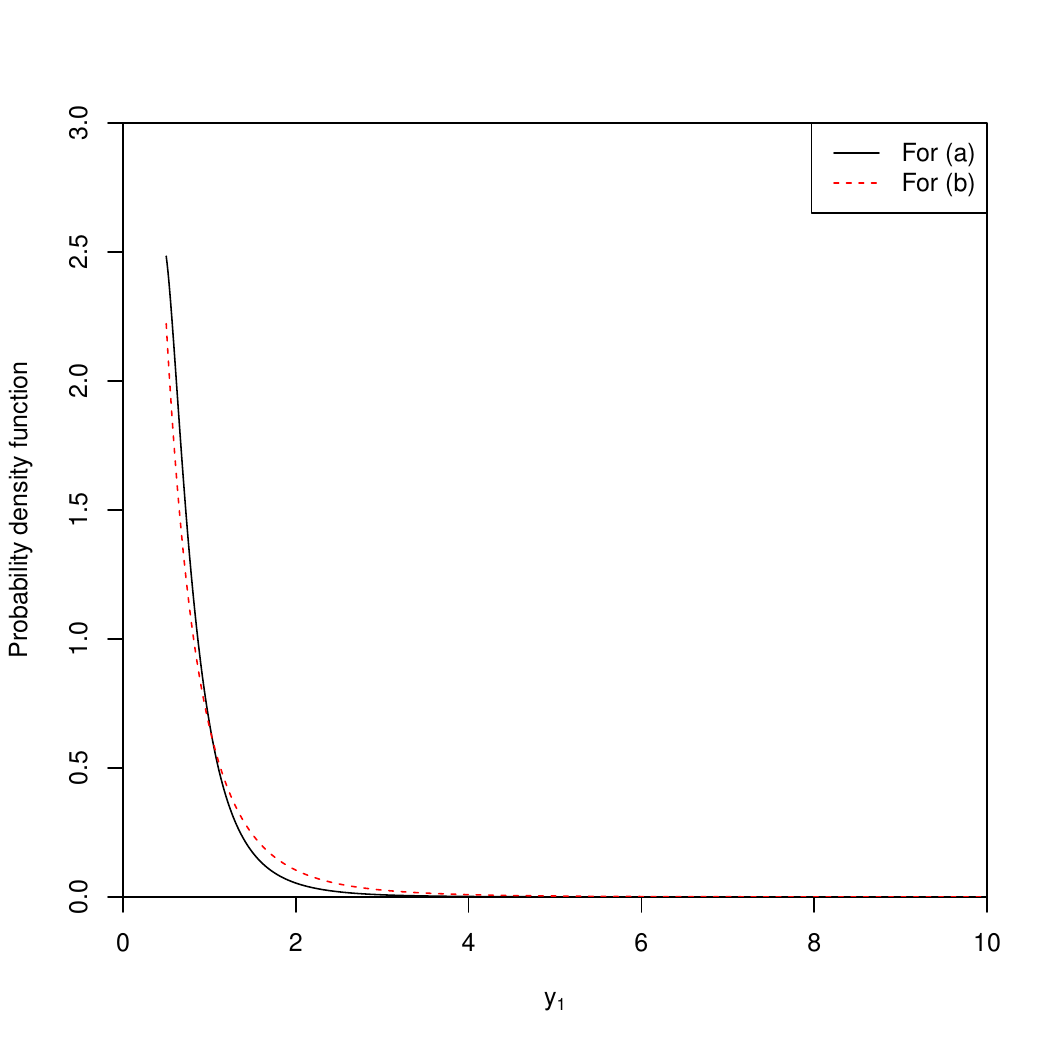}}
	\subfigure[Hazard functions.]{\label{hazard}
		\includegraphics[height=70mm, width=70mm]{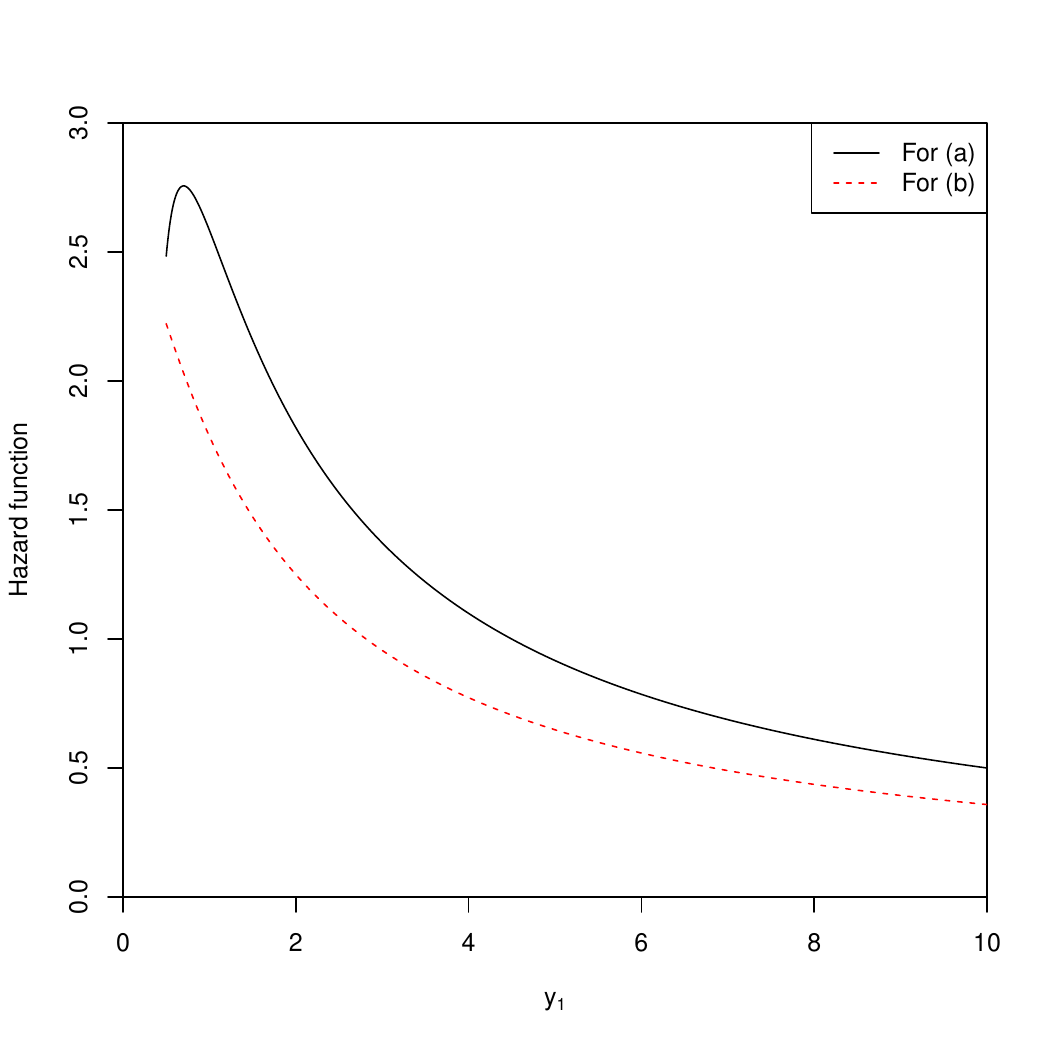}}
	\caption{Probability and hazard functions for (a): $\mu_1 = 0.5, \sigma_1 = 1.5, \alpha_0 = 2.75, \alpha_1 = 2.75, \alpha_2 = 5.0$ and (b): $\mu_1 = 0.5, \sigma_1 = 1.5, \alpha_0 = 2.0, \alpha_1 = 2.0, \alpha_2 = 1.0$.}
	\label{pdfhf}
\end{figure}

\begin{figure}[h]
	\centering
	\subfigure[$\mu_1 = 0, \mu_2 = 0, \sigma_1 = 1, \sigma_2 = 0.5, \alpha_0 = 1, \alpha_1 = 0.3, \alpha_2 = 1.4$]{\label{fig:a}
		\includegraphics[height=60mm, width=60mm]{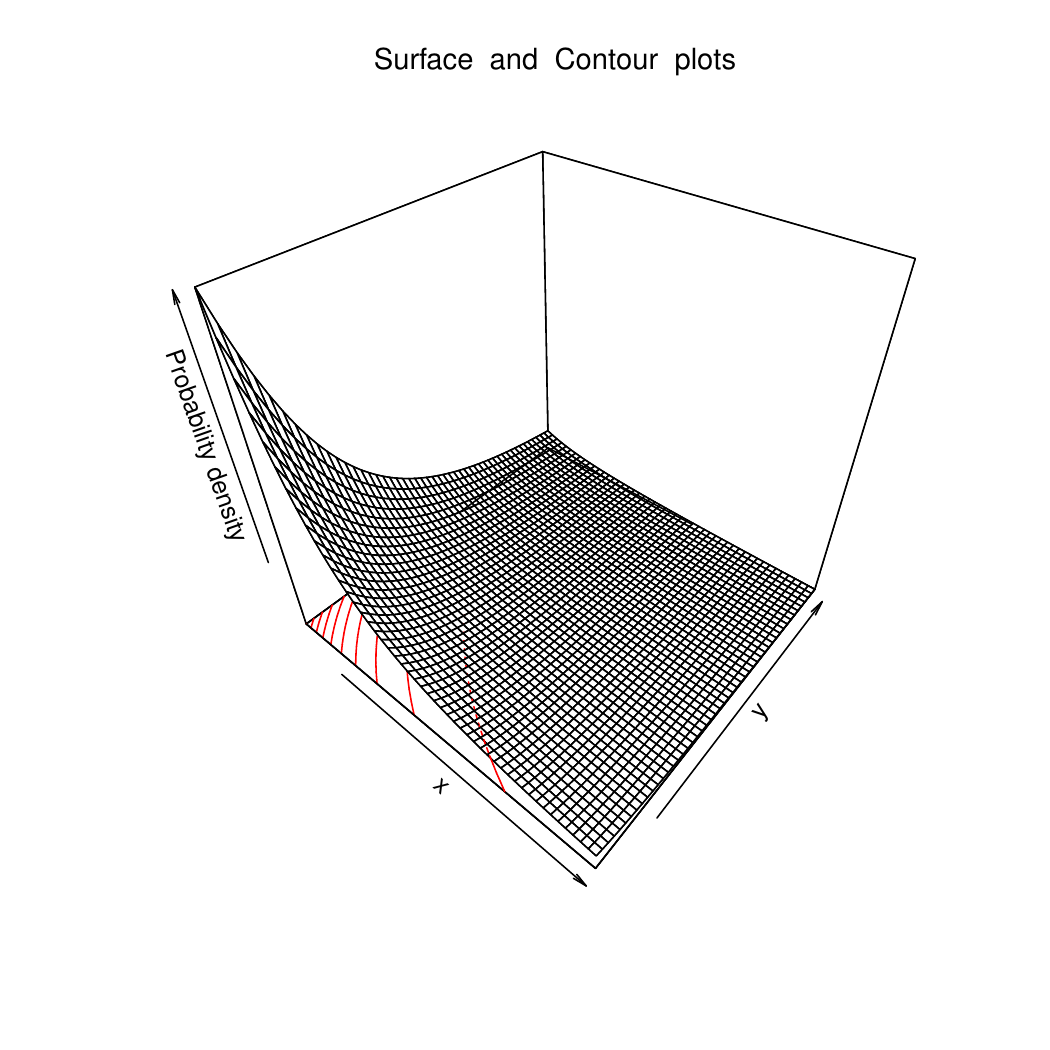}}
	\subfigure[$\mu_1 = 1, \mu_2 = 2, \sigma_1 = 0.4, \sigma_2 = 0.5, \alpha_0 = 2, \alpha_1 = 1.2, \alpha_2 = 1.4$]{\label{fig:b}
		\includegraphics[height=60mm, width=60mm]{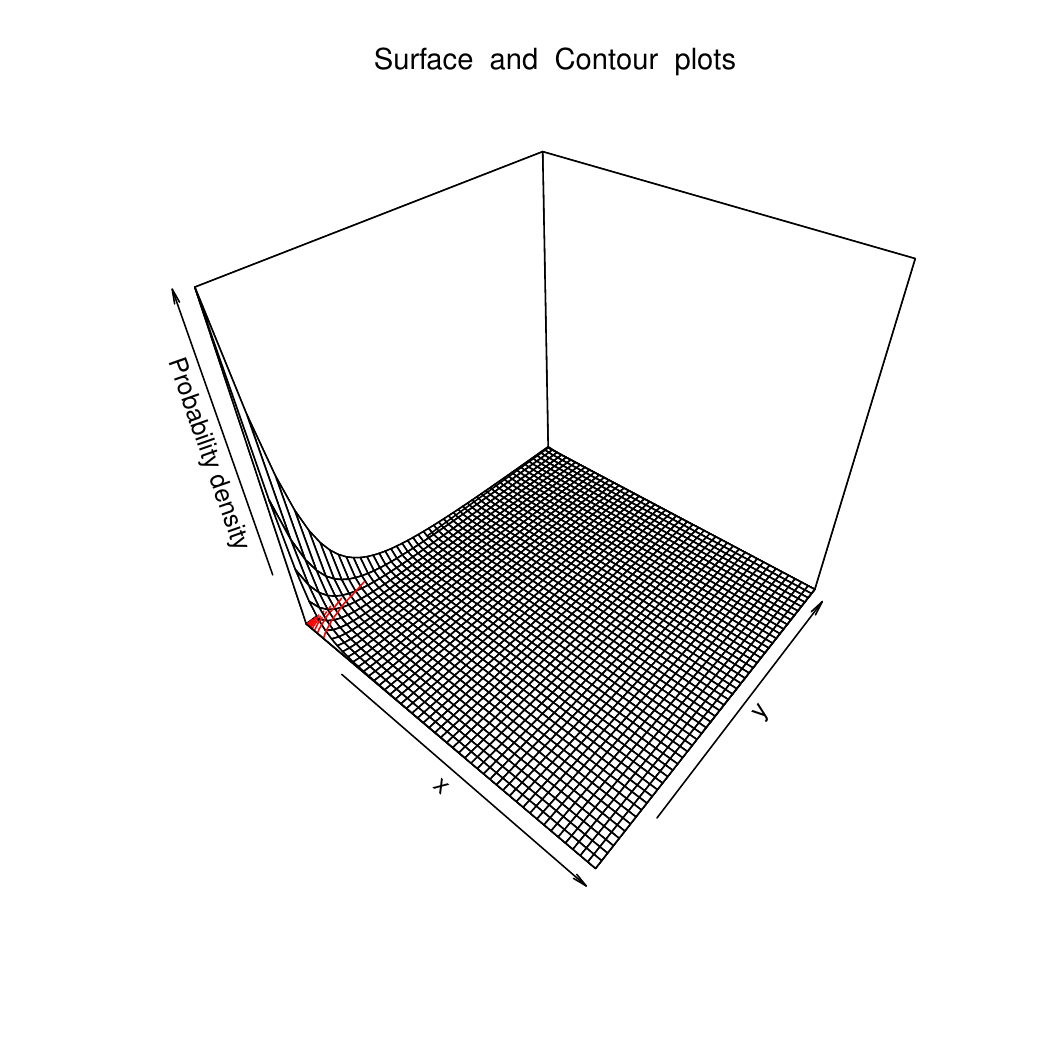}}
	%
	\subfigure[$\mu_1 = 0, \mu_2 = 0, \sigma_1 = 1.4, \sigma_2 = 0.5, \alpha_0 = 1, \alpha_1 = 1, \alpha_2 = 1.4$]{\label{fig:c}
		\includegraphics[height=60mm, width=60mm]{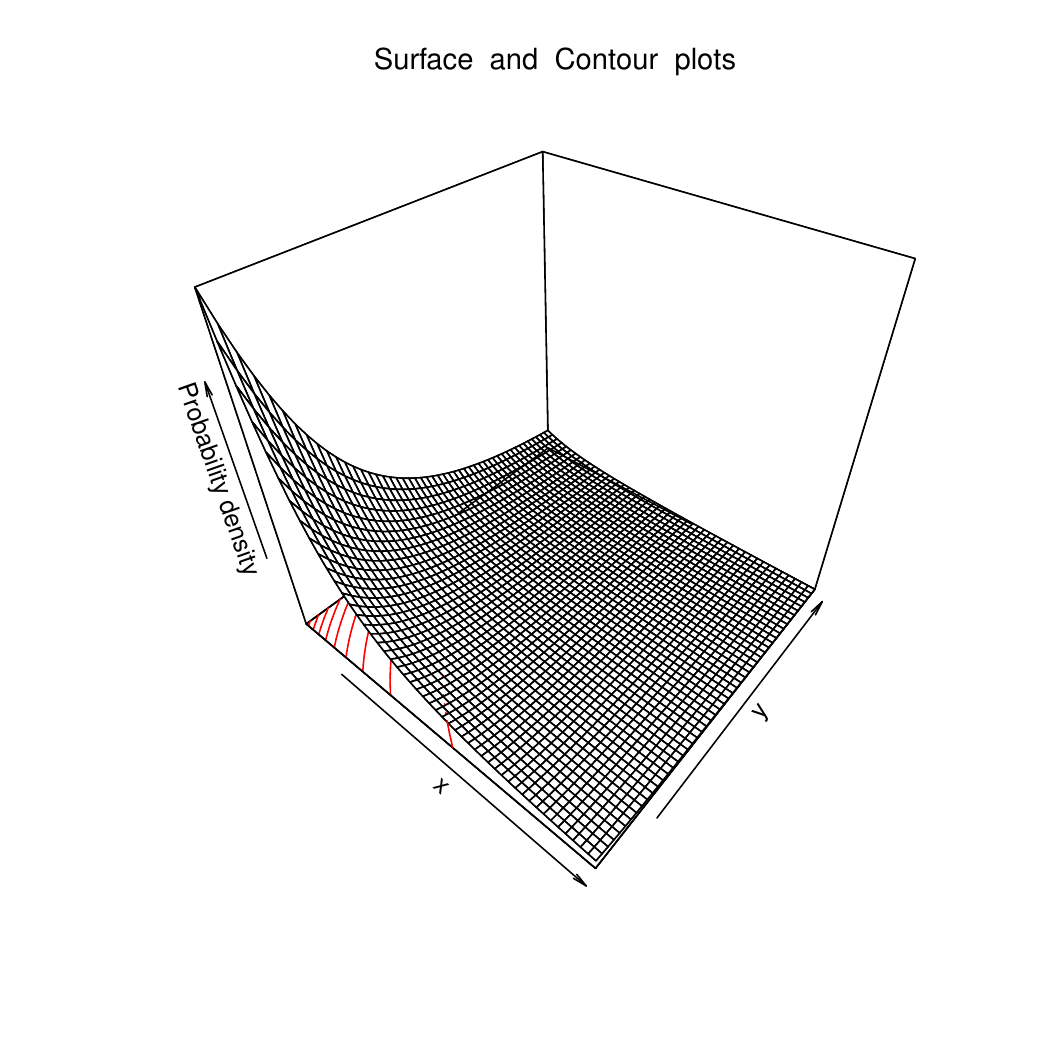}}
	\subfigure[$\mu_1 = 0, \mu_2 = 0, \sigma_1 = 1.4, \sigma_2 = 0.5, \alpha_0 = 2, \alpha_1 = 0.4, \alpha_2 = 0.5$]{\label{fig:d}
		\includegraphics[height=60mm, width=60mm]{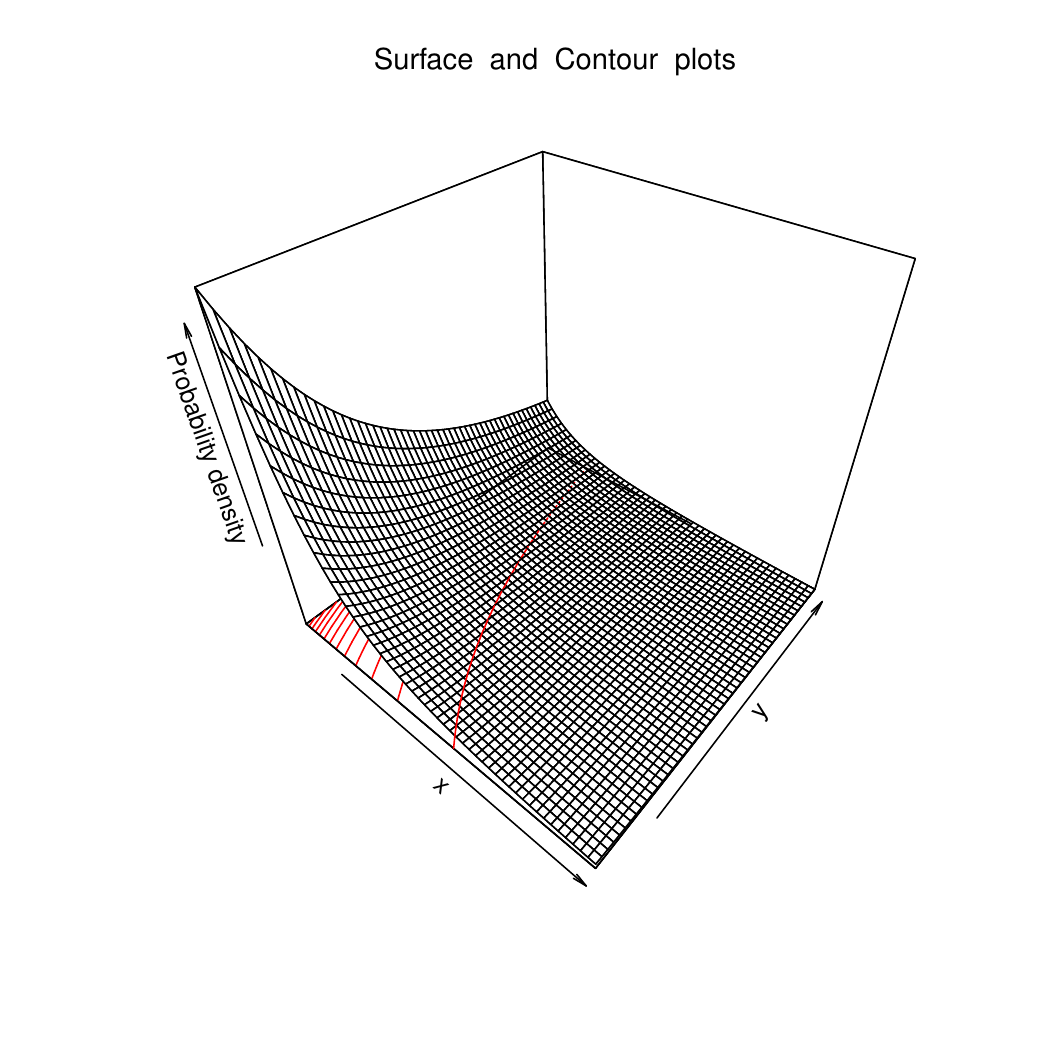}}
	\caption{Surface and contour plots of PDF of BB-BVPA for different values of parameters.}
	\label{fig1_ch3}
\end{figure}

Now we mainly discuss some basic properties of marginal $Y_1$. The properties of $Y_2$ are exactly the same. The marginal survival functions of $Y_1$ is
\begin{eqnarray*}
   S_{Y_{1}}(y_{1}) = cS_{PA}(y_{1}; \mu_{1}, \sigma_{1}, \alpha_{0} + \alpha_{1}) - \frac{\alpha_{0}}{\alpha_{1} + \alpha_{2}}S_{PA}(y_{1}; \mu_{1}, \sigma_{1}, \alpha_{0} + \alpha_{1} + \alpha_{2}).
   \label{surv1}
\end{eqnarray*}
The hazard function (HF) of $Y_1$ is the following: 
\begin{eqnarray*}
	h_{Y_{1}}(y_{1}) = \frac{(\alpha_{0}+\alpha_{1}+\alpha_{2})(1+\frac{y_1-\mu_1}{\sigma_{1}})^{-1}}{\sigma_{1}}\Bigg[1-\frac{\alpha_{2}}{(\alpha_{0}+\alpha_{1}+\alpha_{2}) - \alpha_{0}(1+\frac{y_1-\mu_1}{\sigma_{1}})^{-\alpha_{2}}}\Bigg].\\  
	\label{hazy1}
\end{eqnarray*}
The PDFs and HFs of $Y_1$ for different values of model parameters are shown in Figure~\ref{pdfhf}. Although the hazard function of Pareto type $II$ is decreasing, interestingly, for some values of model parameters, this function first increases and then decreases. For $\mu_1 \geq 0$, the distribution of marginal $Y_1$ can be written as a weighted distribution corresponding to $PA(II)(\mu_1, \sigma_1, \alpha_{0}+\alpha_{1})$ with weight function proportional to 
$$
	w_1(y_1) = 1 - \frac{\alpha_{0}}{\alpha_0+\alpha_1}\bigg(1+\frac{y_1-\mu_1}{\sigma_1}\bigg)^{-\alpha_2},
$$
which is non-negative function with finite non-zero expectation. Hence, the properties available for the weighted distributions can also be applied here.
%
\par From Theorem~\ref{th1}, it is clear that the distribution of $Y_1$ is not  Pareto type $II$ distribution in general. Although the PDF of $Y_1$ is not Pareto type $II$ in general, the shape of the PDF of $Y_1$ is very similar to the PDF of a Pareto type $II$ distribution. Now we discuss the conditional probability density functions. 
\begin{theorem}
	\label{th2}
		If $(Y_1, Y_2)$ $\sim$ $BBBVPA(\mu_1, \mu_2, \sigma_1, \sigma_2, \alpha_0, \alpha_1, \alpha_2)$, then the conditional PDFs of $Y_1|Y_2=y_2$ and $Y_2|Y_1=y_1$ are as follows
	\begin{eqnarray}
		f_{Y_{1}|Y_2}(y_{1}|y_2) = 	
		\begin{cases}
			\frac{(\alpha_0+\alpha_2)f_{PA}(y_1;\mu_1,\sigma_1,\alpha_1)}{\alpha_2+\alpha_0\big\{1-\(1+\frac{y_2-\mu_2}{\sigma_2}\)^{-\alpha_1}\big\}},~ \text{if $\frac{y_1-\mu_1}{\sigma_1}$ \textless $\frac{y_2-\mu_2}{\sigma_2}$}
			\\
			\frac{\alpha_2f_{PA}(y_1;\mu_1,\sigma_1,\alpha_0+\alpha_1)}{\(1+\frac{y_2-\mu_2}{\sigma_2}\)^{-\alpha_0}\big[\alpha_2+\alpha_0\big\{1-\(1+\frac{y_2-\mu_2}{\sigma_2}\)^{-\alpha_1}\big\}\big]},~ \text{if $\frac{y_1-\mu_1}{\sigma_1}$ \textgreater $\frac{y_2-\mu_2}{\sigma_2}$}
		\end{cases} 		
		\label{cond1}
	\end{eqnarray}
and
	\begin{eqnarray}
		f_{Y_{2}|Y_1}(y_{2}|y_1) =  
		\begin{cases}
			\frac{\alpha_1f_{PA}(y_2;\mu_2,\sigma_2,\alpha_0+\alpha_2)}{\(1+\frac{y_1-\mu_1}{\sigma_1}\)^{-\alpha_0}\big[\alpha_1+\alpha_0\big\{1-\(1+\frac{y_1-\mu_1}{\sigma_1}\)^{-\alpha_2}\big\}\big]},~ \text{if $\frac{y_1-\mu_1}{\sigma_1}$ \textless $\frac{y_2-\mu_2}{\sigma_2}$}
			\\
			\frac{(\alpha_0+\alpha_1)f_{PA}(y_2;\mu_2,\sigma_2,\alpha_2)}{\alpha_1+\alpha_0\big\{1-\(1+\frac{y_1-\mu_1}{\sigma_1}\)^{-\alpha_2}\big\}},~ \text{if $\frac{y_1-\mu_1}{\sigma_1}$ \textgreater $\frac{y_2-\mu_2}{\sigma_2}$},
		\end{cases} 		
		\label{cond2}
	\end{eqnarray}
	respectively.
\end{theorem}
\begin{proof}
	The conditional PDFs can be obtained by routine calculation.
\end{proof}
The joint survival function of BB-BVPA distribution is
\begin{equation}
	\label{2e6}
	S_{BB}(y_{1}, y_{2}) =
	\begin{cases}
		S_1(y_1,y_2),~\text{if $\frac{y_1-\mu_1}{\sigma_1}$ \textless $\frac{y_2-\mu_2}{\sigma_2}$}
		\\
		S_2(y_1,y_2),~ \text{if $\frac{y_1-\mu_1}{\sigma_1}$ \textgreater $\frac{y_2-\mu_2}{\sigma_2}$},
	\end{cases} 
\end{equation}
where 
\begin{equation*}
	\begin{split}
		S_1(y_1,y_2)&=cS_{PA}(y_{1};\mu_{1}, \sigma_{1}, \alpha_{1})S_{PA}(y_{2};\mu_{2}, \sigma_{2}, \alpha_{0}+\alpha_{2}) - \frac{\alpha_{0}}{\alpha_{1} + \alpha_{2}}S_{PA}(y_{2}; \mu_{2}, \sigma_{2}, \alpha_{0} + \alpha_{1} + \alpha_{2})\\
		S_2(y_1,y_2)&=cS_{PA}(y_{1};\mu_{1}, \sigma_{1}, \alpha_{0}+\alpha_{1})S_{PA}(y_{2};\mu_{2}, \sigma_{2}, \alpha_{2}) - \frac{\alpha_{0}}{\alpha_{1} + \alpha_{2}}S_{PA}(y_{1}; \mu_{1}, \sigma_{1}, \alpha_{0} + \alpha_{1} + \alpha_{2}).
	\end{split}
\end{equation*}	
We provide the bivariate hazard rate of the BB-BVPA distribution. Note that there are several ways of defining the bivariate 	hazard rates. The bavariate failure rate as given in \cite{basu1971bivariate} of this BB-BVPA distribution is
	\begin{eqnarray*}
		\label{2e7}
		h_{BB}(y_{1}, y_{2}) & = & \frac{f_{BB}(y_1,y_2)}{S_{BB}(y_1, y_2)}\\ & = &
		\begin{cases}
			\frac{(\alpha_{0}+\alpha_{1}+\alpha_{2})\alpha_{1}(\alpha_{0}+\alpha_{2})(1+\frac{y_1-\mu_1}{\sigma_1})^{-\alpha_{1}-1}(1+\frac{y_2-\mu_2}{\sigma_2})^{-1}}{\sigma_{1}\sigma_{2}\{(\alpha_{0}+\alpha_{1}+\alpha_{2})(1+\frac{y_1-\mu_1}{\sigma_{1}})^{-\alpha_{1}} - \alpha_{0}(1+\frac{y_2-\mu_2}{\sigma_2})^{-\alpha_{1}}\}},~\text{if $\frac{y_1-\mu_1}{\sigma_1}$ \textless $\frac{y_2-\mu_2}{\sigma_2}$}
			\\
			\frac{(\alpha_{0}+\alpha_{1}+\alpha_{2})\alpha_{2}(\alpha_{0}+\alpha_{1})(1+\frac{y_1-\mu_1}{\sigma_1})^{-1}(1+\frac{y_2-\mu_2}{\sigma_2})^{-\alpha_{2}-1}}{\sigma_{1}\sigma_{2}\{(\alpha_{0}+\alpha_{1}+\alpha_{2})(1+\frac{y_2-\mu_2}{\sigma_{2}})^{-\alpha_{2}} - \alpha_{0}(1+\frac{y_1-\mu_1}{\sigma_1})^{-\alpha_{2}}\}},~ \text{if $\frac{y_1-\mu_1}{\sigma_1}$ \textgreater $\frac{y_2-\mu_2}{\sigma_2}$}.
		\end{cases} 
	\end{eqnarray*}

The hazard gradients \cite[see, e.g.,][]{johnson1975vector} of BB-BVPA model can be obtained as
\begin{eqnarray*}
	\label{2e8}
	h_{1}(y_{1}, y_{2}) & = & - \pdv{}{y_1}S_{BB}(y_1, y_2) \\ & = &
	\begin{cases}
	    g_{1}(y_1, y_2),~\text{if $\frac{y_1-\mu_1}{\sigma_1}$ \textless $\frac{y_2-\mu_2}{\sigma_2}$}
		\\
	   g_{2}(y_1, y_2),~ \text{if $\frac{y_1-\mu_1}{\sigma_1}$ \textgreater $\frac{y_2-\mu_2}{\sigma_2}$},
	\end{cases} 
\end{eqnarray*}
where 
\begin{equation*}
	\begin{split}
		g_1(y_1,y_2)&=cf_{PA}(y_{1};\mu_{1}, \sigma_{1}, \alpha_{1})S_{PA}(y_{2};\mu_{2}, \sigma_{2}, \alpha_{0}+\alpha_{2})\\
		g_2(y_1,y_2)&=cf_{PA}(y_{1};\mu_{1}, \sigma_{1}, \alpha_{0}+\alpha_{1})S_{PA}(y_{2};\mu_{2}, \sigma_{2}, \alpha_{2}) - \frac{\alpha_{0}}{\alpha_{1} + \alpha_{2}}f_{PA}(y_{1}; \mu_{1}, \sigma_{1}, \alpha_{0} + \alpha_{1} + \alpha_{2}).
	\end{split}
\end{equation*}	
and
\begin{eqnarray*}
	\label{2e9}
	h_{2}(y_{1}, y_{2}) & = & - \pdv{}{y_2}S_{BB}(y_1, y_2) \\ & = &
	\begin{cases}
		g'_{1}(y_1, y_2),~\text{if $\frac{y_1-\mu_1}{\sigma_1}$ \textless $\frac{y_2-\mu_2}{\sigma_2}$}
		\\
		g'_{2}(y_1, y_2),~ \text{if $\frac{y_1-\mu_1}{\sigma_1}$ \textgreater $\frac{y_2-\mu_2}{\sigma_2}$},
	\end{cases} 
\end{eqnarray*}
where
\begin{equation*}
	\begin{split}
		g'_1(y_1,y_2)&=cS_{PA}(y_{1};\mu_{1}, \sigma_{1}, \alpha_{1})f_{PA}(y_{2};\mu_{2}, \sigma_{2}, \alpha_{0}+\alpha_{2}) - \frac{\alpha_{0}}{\alpha_{1} + \alpha_{2}}f_{PA}(y_{2}; \mu_{2}, \sigma_{2}, \alpha_{0} + \alpha_{1} + \alpha_{2})\\
		g'_2(y_1,y_2)&=cS_{PA}(y_{1};\mu_{1}, \sigma_{1}, \alpha_{0}+\alpha_{1})f_{PA}(y_{2};\mu_{2}, \sigma_{2}, \alpha_{2}).
	\end{split}
\end{equation*}
%
\begin{theorem}
	\label{th3}
	If $(Y_1, Y_2)$ $\sim$ $BBBVPA(\alpha_0, \alpha_1, \alpha_2)$, then we have the following,
	\begin{enumerate}[(a)] 
		\item The stress-strength parameter $R = P(Y_ 1< Y_2) =  	\frac{\alpha_1}{\alpha_1 + \alpha_2}$,\\ 
		\item $\min\{Y_1, Y_2\}$ $\sim PA(II)(0, 1, \alpha_0+\alpha_1 + \alpha_2)$,\\
		\item $Y_1|\{Y_1 < Y_2\}$ $\sim PA(II)(0, 1, \alpha_0+\alpha_1 + \alpha_2)$,\\
		\item $Y_2|\{Y_2 < Y_1\}$ $\sim PA(II)(0, 1, \alpha_0+\alpha_1 + \alpha_2)$.
	\end{enumerate}
\end{theorem}
\begin{proof}
	The proofs are quite trivial and, therefore, omitted.
\end{proof}
%
%
Now we provide the total positivity result of $(Y_1, Y_2)$, for identical marginals.
\begin{theorem}
	\label{th6}
	If $(Y_1, Y_2)$ $\sim$ $BBBVPA(\alpha_0, \alpha_1, \alpha_2)$, then the joint PDF of $(Y_1, Y_2)$ has $TP_2$ property.
\end{theorem}
\begin{proof}
	Note that $(Y_1, Y_2)$ has $TP_2$ \cite[see, e.g.,][]{karlin1968total} property, iff for any $y_{11}$, $y_{12}$, $y_{21}$, $y_{22}$, whenever $0 < y_{11} < y_{12}$, and $0 < y_{21} < y_{22}$, we have 
	\begin{eqnarray}
		\label{tp2}
		f_{BB}(y_{11}, y_{21})f_{BB}(y_{12}, y_{22}) \geq f_{BB}(y_{12}, y_{21})f_{BB}(y_{11}, y_{22}).
	\end{eqnarray}
In our case, we have six situations as follows: $y_{21}<y_{22}<y_{11}<y_{12}$, $y_{21}<y_{11}<y_{22}<y_{12}$, $y_{21}<y_{11}<y_{12}<y_{22}$, $y_{11}<y_{12}<y_{21}<y_{22}$, $y_{11}<y_{21}<y_{22}<y_{12}$ and $y_{11}<y_{21}<y_{12}<y_{22}$.  Now, different cases are considered as follows: \\ 
\textit{Case 1:} when $y_{21}<y_{22}<y_{11}<y_{12}$, we have
	\begin{eqnarray*}
	& & f_{BB}(y_{11}, y_{21})f_{BB}(y_{12}, y_{22}) - f_{BB}(y_{12}, y_{21})f_{BB}(y_{11}, y_{22}) = 0
	.
\end{eqnarray*}
\\
\textit{Case 2:} $y_{21}<y_{11}<y_{22}<y_{12}$, \\ By simple calculation, it can be shown that to prove \eqref{tp2} is equivalent to prove
$$ (1+y_{11})^{-\alpha_{0}} \geq (1+y_{22})^{-\alpha_{0}},$$
which is true in this case as $y_{11}<y_{22}$. Similarly, for other cases also, it can be proved in the similar way.
\end{proof}
It is a very strong type of dependence in the sense that it implies most of the other types of dependence. This type of dependence is also known as positive likelihood ratio dependence. 
\par As we mentioned that the log-likelihood function for seven-parameter BB-BVPA distribution is discontinuous with respect to location and scale parameters, only the three-parameter BB-BVPA model satisfies all the regularity conditions for the MLEs to be consistent and asymptotically normal and we can state the following result:
\begin{theorem}
	\label{th5}
	If $\hat{\alpha}_0$, $\hat{\alpha}_1$ and $\hat{\alpha}_2$ are the MLEs of the parameters $\alpha_0$, $\alpha_1$ and $\alpha_2$, respectively, of three-parameter BB-BVPA distribution, then
	$$\sqrt{n}\big[\hat{\alpha}_0-\alpha_0, \hat{\alpha}_1-\alpha_1, \hat{\alpha}_2-\alpha_2 \big] \sim N_3\big(0, I^{-1}\big), $$
	here $I$ is the Fisher information matrix.
\end{theorem}
\begin{proof}
	The proof is quite obvious, therefore, omitted.
\end{proof}
\section{EM algorithm for Block-Basu bivariate Pareto distribution}
\label{s3}
This section addresses the problem of computing the estimators through the EM algorithm of both three and seven parameters BB-BVPA distributions.

We start by discussing the EM algorithm for three-parameter MOBVPA distribution. Let us assume $\mathcal{I} = \{(x_{11},x_{21}),(x_{12},x_{22}),\cdots,(x_{1n},x_{2n})\}$ is a random sample of size $n$ from $MOBVPA(\alpha_0, \alpha_1,$ $ \alpha_2)$. Let us define the following notations: $ I_0 = \{ i : x_{1i} = x_{2i} \}$, $ I_1 = \{ i : x_{1i}  < x_{2i} \}$, $ I_2 = \{ i : x_{1i} > x_{2i} \} $ and $n_0 = |I_0|$, $n_1 = |I_1|$, $n_2 = |I_2|$. As expected MLEs of $\alpha_0$, $\alpha_1$, and $\alpha_2$ cannot be obtained in explicit form. To compute the MLEs directly, one needs to solve a three-dimensional optimization problem. If all the $U_0$, $U_1$ and $U_2$ are known, the MLEs of the unknown parameters can be obtained by solving a one-dimensional optimization problem. We treat this problem as a missing value problem and propose to use the EM algorithm for computing the MLEs of unknown parameters. Usual EM implementation requires the identification of some missing structure within the problem. Here we do not know whether $X_1$ is $U_0$ or $U_1$ and similarly we also have no information regarding $X_2$ whether it is $U_0$ or $U_2$. Let us introduce a pair of random variables $(\Delta_1, \Delta_2)$ associated with each $(X_1,X_2)$ as,
\begin{equation*}
	\Delta_1 = 
	\begin{cases}  
		0  & \text{if} ~ X_1 = U_{0} \\ 1  & \text{if} ~ X_1 = U_1 \\  
	\end{cases} \qquad
	\Delta_2 = 
	\begin{cases}
		0  & \text{if} ~ X_2 = U_{0}  \\ 2  & \text{if} ~ X_2 = U_2.
	\end{cases}
\end{equation*}
In this case, even if we know $(X_1, X_2)$, the corresponding $(\Delta_1, \Delta_2)$ may not always be known. For example, if
$X_1 \ne X_2$, then $(\Delta_1, \Delta_2)$ is not known; but it is $(0,0)$ when $X_1=X_2$. If $i \in I_1$, i.e., $ x_{1i} < x_{2i}$, then the possible values of $(\Delta_1, \Delta_2)$ are $(1, 0)$ or $(1, 2)$ with non-zero probabilities $ u_1 = P(\Delta_2= 0|I_1 )$ or $ u_2 = P(\Delta_2= 2|I_1 )$, and similarly,  if $i\in I_2$, i.e., $ x_{1i} > x_{2i}$,  then the possible values of $(\Delta_1, \Delta_2)$ are $(0, 2)$ or $(1, 2)$ with non-zero probabilities $ w_1 = P(\Delta_1= 0|I_2)$ or $ w_2 = P(\Delta_1= 2|I_2)$.

Note that if $(X_1, X_2)$ and the associated $(\Delta_1, \Delta_2)$ are known for all the observations, then the MLEs of the unknown parameters $\alpha_0$, $\alpha_1$ and $\alpha_2$ can be obtained very easily, by solving a one-dimensional optimization problem. But unfortunately $(\Delta_1, \Delta_2)$ are not known for all the observations. To implement the EM algorithm, first we obtain the E-step \cite[see, e.g.,][]{dinse1982nonparametric}. In E-step the `pseudo-log-likelihood' function is formed from the log-likelihood function by replacing the log-likelihood contribution of  $(X_1, X_2)$ by its expected value, if the corresponding $(\Delta_1, \Delta_2)$ is missing. In M-step, we estimate the unknown parameters by maximizing the `pseudo-log-likelihood' function with respect to the unknown parameters. The observations in Table~\ref{tab1} are used for constructing E-step.
%
\begin{table}[H]
	\begin{center}
	  \begin{adjustbox}{width=0.34\textwidth}
		\begin{tabular}{|c|c|r|}
			\hline
			Ordering &  $(X_1,X_2)$   & Group  \\
			\hline
			$U_0 < U_1 < U_2$  & $(U_0,U_0)$ & $I_0$\\
			$U_0 < U_2 < U_1$  & $(U_0,U_0)$ & $I_0$\\
			$U_1 < U_0 < U_2$  & $(U_1,U_0)$ & $I_1$\\
			$U_1 < U_2 < U_0$  & $(U_1,U_2)$ & $I_1$\\
			$U_2 < U_0 < U_1$  & $(U_0,U_2)$ & $I_2$\\
			$U_2 < U_1 < U_0$  & $(U_1,U_2)$ & $I_2$\\
			\hline
		\end{tabular}
		\end{adjustbox}
		\caption{Groups and corresponding orderings of hidden random variables $U_0$, $U_{1}$ and $U_{2}$. \label{tab1}}  
	\end{center}
\end{table}
Since 
\begin{align*}
	P(U_1 < U_0 < U_2) =\frac{\alpha_0 \alpha_1}{(\alpha_0 + \alpha_2)(\alpha_0 +\alpha_1 + \alpha_2)},
\end{align*}
and 
\begin{align*}
	P(U_1 < U_2 < U_0) = \frac{\alpha_1 \alpha_2}{(\alpha_0 + \alpha_2)(\alpha_0 +\alpha_1 + \alpha_2)},
\end{align*}
we have the following expressions for $u_1$ and $u_2$,
$$u_1=\frac{P(U_1 < U_0 < U_2)}{P(U_1 < U_0 < U_2)+P(U_1 < U_2 < U_0)} = \frac{\alpha_0}{\alpha_0 +\alpha_2}$$ and
$$u_2=\frac{P(U_1 < U_2 < U_0)}{P(U_1 < U_0 < U_2)+P(U_1 < U_2 < U_0)} = \frac{\alpha_2}{\alpha_0 +\alpha_2}.$$
Similarly, we can calculate
$$w_1=\frac{P(U_2 < U_0 < U_1)}{P(U_2 < U_0 < U_1)+P(U_2 < U_1 < U_0)} = \frac{\alpha_0}{\alpha_0 +\alpha_1}$$ and
$$w_2=\frac{P(U_2 < U_1 < U_0)}{P(U_2 < U_0 < U_1)+P(U_2 < U_1 < U_0)} = \frac{\alpha_1}{\alpha_0 +\alpha_1}.$$
Therefore, the pseudo-log-likelihood function can be written as,
\begin{equation}
		\label{pse-ld}
	\begin{split}
			L^{*}(\alpha_{0}, \alpha_{1}, \alpha_{2})  = & -\alpha_0 \bigg(\sum\limits_{i\in I_0} \ln (1 + x_i) + \sum\limits_{i\in I_2} \ln (1 + x_{1i}) + \sum\limits_{i\in I_1} \ln (1 + x_{2i})\bigg) \\&  + (n_0 + u_1n_1 + w_1n_2) \ln \alpha_0  - \alpha_1\bigg(\sum\limits_{i\in I_0} \ln (1 + x_i) + \sum\limits_{i\in I_1 \cup I_2} \ln (1 + x_{1i})\bigg)\\& + (n_1 + w_2n_2)\ln \alpha_1  - \alpha_2\bigg(\sum\limits_{i\in I_0} \ln (1 + x_i) + \sum\limits_{i\in I_1\cup I_2} \ln (1 + x_{2i})\bigg) \\&  + (n_2 + u_2n_1)\ln \alpha_2. 
	\end{split}
\end{equation}
M-step involves maximizing \eqref{pse-ld} with respect to (w.r.t.) $\alpha_0$, $\alpha_1$ and $\alpha_2$. The maximization of \eqref{pse-ld} w.r.t. $\alpha_0$, $\alpha_1$ and $\alpha_2$ can be obtained at 
\begin{equation} 
	\hat{\alpha}_{0} = \frac{n_{0} + u_{1} n_{1} + w_{1} n_{2}}{\sum_{i \in I_{0}}^{} \ln(1 + x_{i}) + \sum_{i \in I_{2}}^{} \ln(1 + x_{1i}) + \sum_{i \in I_{1}}^{} \ln (1 + x_{2i})}, \label{alph0}
\end{equation}
\begin{equation}
	 \hat{\alpha}_{1} = \frac{(n_{1} + w_{2}n_{2})}{\sum_{i \in I_{0}}^{} \ln(1 + x_{i})  + \sum_{i \in (I_{1} \cup I_{2})}^{} \ln(1 + x_{1i}) }, \label{alph1}
 \end{equation}
\begin{equation} 
	\hat{\alpha}_{2} = \frac{n_{2} + u_{2} n_{1}}{\sum_{i \in I_{0}}^{} \ln(1 + x_{i}) +  \sum_{i \in (I_{1} \cup I_{2})}^{} \ln(1 + x_{2i})}. \label{alph2}
\end{equation}
The following algorithm describes how to obtain the $(i+1)$th step from
the $i$th step of the EM algorithm. Suppose that at the $i$th step the estimates of $\alpha_0$, $\alpha_1$, $\alpha_2$ are  $\hat{\alpha}^{(i)}_{0}$, $\hat{\alpha}^{(i)}_{1}$ and $\hat{\alpha}^{(i)}_{2}$, respectively.
\begin{algorithm}[H]
	\caption{EM algorithm of three-parameter MOBVPA distribution.	\label{algo1_ch3}}
	\text{Start with some initial choice of parameters $\alpha_0$, $\alpha_1$, $\alpha_2$}.
	\begin{algorithmic}[1]
		\STATE Compute $u^{(i)}_{1}$, $u^{(i)}_{2}$, $w^{(i)}_{1}$, $w^{(i)}_{2}$ using $\hat{\alpha}^{(i)}_{0}$, $\hat{\alpha}^{(i)}_{1}$, $\hat{\alpha}^{(i)}_{2}$.
		\STATE Update $\hat{\alpha}^{(i+1)}_{0}$, $\hat{\alpha}^{(i+1)}_{1}$ and $\hat{\alpha}^{(i+1)}_{2}$ using equations \eqref{alph0}, \eqref{alph1} and \eqref{alph2}, respectively.
	\end{algorithmic}
    \text{The process should be continued until the convergence criterion is met. In our work, the process} \text{is stopped when $\frac{\abs{L^{*}\(\hat{\alpha}^{(i+1)}_{0}, \hat{\alpha}^{(i+1)}_{1}, \hat{\alpha}^{(i+1)}_{2}\) - L^{*}\(\hat{\alpha}^{(i)}_{0}, \hat{\alpha}^{(i)}_{1}, \hat{\alpha}^{(i)}_{2}\)}}{L^{*}\(\hat{\alpha}^{(i)}_{0}, \hat{\alpha}^{(i)}_{1}, \hat{\alpha}^{(i)}_{2}\) } < \epsilon$, where $\epsilon = 10^{-5}$.}
\end{algorithm}


\subsection{EM algorithm for three-parameter BB-BVPA distribution}
Let us consider $ \mathcal{D}_1 = \{(y_{11},y_{21}),(y_{12},y_{22}),\cdots,(y_{1n},y_{2n})\}$ is a random sample of size $n$ from three-parameter BB-BVPA distribution. We use the following notation;
\begin{center}
	$D_1 = \{i \mid y_{1i} < y_{2i}\}$, $D_2 = \{i \mid y_{1i} > y_{2i}\}$, 	$|D_1| = n_1$, 	$|D_2| = n_2$, and $n=n_1+n_2$,
\end{center}
where $|D_j|$ for $j = 1, 2$ denotes the number of elements in the set $D_j$. The log-likelihood function based on $\mathcal{D}_1$ is
\begin{equation*}
	\label{3e1}
	\begin{split}
	   L(\alpha_{0}, \alpha_{1}, \alpha_{2})  = &~(n_{1} + n_{2})\ln (\alpha_{0} + \alpha_{1} + \alpha_{2}) -  (n_{1} + n_{2})\ln(\alpha_{1} + \alpha_{2}) + \sum_{i \in D_{1}}^{} \ln f_{PA}(y_{1i}; 0, 1, \alpha_{1}) \\  + &  \sum_{i \in D_{1}}^{} \ln f_{PA}(y_{2i}; 0, 1, \alpha_{0} + \alpha_{2}) + \sum_{i \in D_{2}}^{} \ln f_{PA}(y_{1i}; 0, 1, \alpha_{0} + \alpha_{1}) \\  + & \sum_{i \in D_{2}}^{} \ln f_{PA}(y_{2i}; 0, 1, \alpha_{2})\\
	   = &~ (n_{1} + n_{2})\ln (\alpha_{0} + \alpha_{1} + \alpha_{2}) - (n_{1} + n_{2})\ln (\alpha_{1} + \alpha_{2}) +  n_{1}\ln\alpha_{1} \\  - & (\alpha_{1} + 1)\sum_{i \in D_{1}}^{} \ln(1 + y_{1i}) + n_{1}\ln(\alpha_{0} + \alpha_{2}) - (\alpha_{0} + \alpha_{2} + 1)\sum_{i \in D_{1}}^{} \ln(1 + y_{2i})\\  + &  n_{2}\ln(\alpha_{0} + \alpha_{1}) - (\alpha_{0} + \alpha_{1} + 1)\sum_{i \in D_{2}}^{} \ln(1 + y_{1i}) + n_{2}\ln\alpha_{2} - (\alpha_{2} + 1)\sum_{i \in D_{2}}^{} \ln(1 + y_{2i}).
	\end{split} 
\end{equation*}
%
In this three-parameter BB-BVPA case, the whole set $I_0$ (described above) is considered as the missing observations. We adopt similar existing methods described in \cite{kundu2010class}, and replace the missing quantity $n_{0}$ (cardinality of $I_0$) and each observation of $U_{0}$ falling within $I_{0}$ by its estimate $\tilde{n}_{0}$ and $ a = E(U_{0}| U_{0} < \min\{ U_{1}, U_{2}\})$, respectively.  Note that in this case, $n_{0}$ is a random variable which has the negative binomial with parameters $(n_{1} + n_{2})$ and $\frac{\alpha_{1} + \alpha_{2}}{\alpha_{0} + \alpha_{1} + \alpha_{2}}$. Therefore, $$ \tilde{n}_{0} = (n_{1} + n_{2})\frac{\alpha_{0}}{\alpha_{1} + \alpha_{2}}~\text{and}~a = E\big[U_{0} | U_{0} < \min\{U_{1}, U_{2}\}\big] = \frac{1}{(\alpha_{0} + \alpha_{1} + \alpha_{2} - 1)}. $$
\textbf{ An important restriction for this approximation is that we have to ensure $\alpha_{0} + \alpha_{1} + \alpha_{2} > 1$. The restriction will ensure the existence of the above expectation.}

Therefore, under the restriction mentioned above, we can write the pseudo-log-likelihood function by replacing the missing observations with their expected value as follows.
\begin{equation}
	\begin{split}
			Q(\alpha_{0}, \alpha_{1}, \alpha_{2})  = ~& -\alpha_0 \bigg(\tilde{n}_{0} \ln (1 + a) + \sum\limits_{i \in D_2} \ln (1 + y_{1i}) + \sum\limits_{i \in D_1} \ln (1 + y_{2i})\bigg) \\  + ~& (\tilde{n}_0 + u_1n_1 + w_1n_2) \ln \alpha_0  - \alpha_1\bigg(\tilde{n}_{0} \ln (1 + a) + \sum\limits_{i\in D_1 \cup D_2} \ln (1 + y_{1i})\bigg) \\ + ~& (n_1 + w_2n_2)\ln \alpha_1  - \alpha_2\bigg(\tilde{n}_{0}\ln (1 + a) + \sum\limits_{i\in D_1\cup D_2} \ln (1 + y_{2i})\bigg) \\  + ~& (n_2 + u_2n_1)\ln \alpha_2. 
	\end{split}
	\label{likelihood1}
\end{equation}
At each step, the pseudo-maximum likelihood estimates of $\alpha_{0}$, $\alpha_{1}$ and $\alpha_{2}$ can be obtained using \eqref{likelihood1} as 
\begin{equation} 
	\hat{\alpha}_{0} = \frac{\tilde{n}_{0} + u_{1} n_{1} + w_{1} n_{2}}{\tilde{n}_{0} \ln(1 + a) + \sum_{i \in D_{2}}^{} \ln(1 + y_{1i}) + \sum_{i \in D_{1}}^{} \ln (1 + y_{2i})}, \label{alph00}
\end{equation}
\begin{equation} 
	\hat{\alpha}_{1} = \frac{n_{1} + w_{2}n_{2}}{\tilde{n}_{0} \ln(1 + a)  + \sum_{i \in D_{1} \cup D_{2}}^{} \ln(1 + y_{1i})}, \label{alph11}
\end{equation}
and
\begin{equation}
	\hat{\alpha}_{2} = \frac{n_{2} + u_{2} n_{1}}{\tilde{n}_{0} \ln(1 + a) +  \sum_{i \in D_{1} \cup D_{2}}^{} \ln(1 + y_{2i})}. \label{alph22}
\end{equation}
Therefore, one can describe how to obtain the $(i+1)$th step from
the $i$th step of the EM algorithm is as follows. 
\begin{algorithm}[H]
	\caption{EM algorithm of three-parameter BB-BVPA distribution.	\label{algo2_ch3}}
	\text{Start with some initial choice of parameters $\alpha_0$, $\alpha_1$, $\alpha_2$}.
	\begin{algorithmic}[1]
		\STATE Compute $u^{(i)}_{1}$, $u^{(i)}_{2}$, $w^{(i)}_{1}$, $w^{(i)}_{2}$, $\tilde{n}_{0}$ and $a$ using $\hat{\alpha}^{(i)}_{0}$, $\hat{\alpha}^{(i)}_{1}$ and $\hat{\alpha}^{(i)}_{2}$.
		\STATE Update $\hat{\alpha}^{(i+1)}_{0}$, $\hat{\alpha}^{(i+1)}_{1}$ and $\hat{\alpha}^{(i+1)}_{2}$ using equations \eqref{alph00}, \eqref{alph11} and \eqref{alph22}, respectively.
	\end{algorithmic}
    \text{The process should be continued until the convergence criterion is met. In our work, the process} \text{is stopped when $\frac{\abs{Q\(\hat{\alpha}^{(i+1)}_{0}, \hat{\alpha}^{(i+1)}_{1}, \hat{\alpha}^{(i+1)}_{2}\) - Q\(\hat{\alpha}^{(i)}_{0}, \hat{\alpha}^{(i)}_{1}, \hat{\alpha}^{(i)}_{2}\)}}{Q\(\hat{\alpha}^{(i)}_{0}, \hat{\alpha}^{(i)}_{1}, \hat{\alpha}^{(i)}_{2}\) } < \epsilon$, where $\epsilon = 10^{-5}$.}
\end{algorithm}

%
\noindent{\textbf{Important remark}} The above method fails when $\alpha_{0} + \alpha_{1} + \alpha_{2} < 1$.  
\paragraph*{Proposed modification}  
To make the algorithm valid for any range of parameters, instead of estimating $U_{0}$, we estimate $\ln (1 + U_{0})$ conditional on $U_{0} < \min\{ U_{1}, U_{2} \}$.  Since $\ln (x)$ is an increasing function of $x$, our condition is equivalent to $\ln(1 + U_{0}) < \min\{ \ln(1 + U_{1}), \ln(1 + U_{2}) \}$.
Therefore, we replace the unknown missing information $\ln(1 + U_{0})$ by $$a^{*} = E\big[\ln(1 + U_{0}) | \ln(1 + U_{0}) < \min\{\ln(1 + U_{1}), \ln(1 + U_{2})\}\big] = \frac{1}{(\alpha_{0} + \alpha_{1} + \alpha_{2})}.$$ Therefore, the pseudo-maximum likelihood estimates of $\alpha_{0}$, $\alpha_{1}$ and $\alpha_{2}$ are

\begin{equation} 
	\hat{\alpha}_{0} = \frac{\tilde{n}_{0} + u_{1} n_{1} + w_{1} n_{2}}{\tilde{n}_{0} a^{*} + \sum_{i \in D_{2}}^{} \ln(1 + y_{1i}) + \sum_{i \in D_{1}}^{} \ln (1 + y_{2i})}, \label{alph000}
\end{equation}
\begin{equation} 
	\hat{\alpha}_{1} = \frac{n_{1} + w_{2}n_{2}}{\tilde{n}_{0} a^{*}  + \sum_{i \in D_{1} \cup D_{2}}^{} \ln(1 + y_{1i})}, \label{alph111}
\end{equation}
\begin{equation} 
	\hat{\alpha}_{2} = \frac{n_{2} + u_{2} n_{1}}{\tilde{n}_{0} a^{*} +  \sum_{i \in D_{1} \cup D_{2}}^{} \ln(1 + y_{2i})}. \label{alph222}
\end{equation}
Therefore, start with some initial choice of parameters $\alpha_0$, $\alpha_1$, $\alpha_2$, the $(i+1)$th step of modified EM algorithm can be written as  
\begin{algorithm}[H]
	\caption{Modified EM algorithm of three-parameter BB-BVPA distribution.	\label{algo3_ch3}}
	\text{Start with some initial choice of parameters $\alpha_0$, $\alpha_1$, $\alpha_2$}.
	\begin{algorithmic}[1]
		\STATE Compute $u^{(i)}_{1}$, $u^{(i)}_{2}$, $w^{(i)}_{1}$, $w^{(i)}_{2}$, $\tilde{n}_{0}$ and $a^*$ using $\hat{\alpha}^{(i)}_{0}$, $\hat{\alpha}^{(i)}_{1}$ and $\hat{\alpha}^{(i)}_{2}$.
		\STATE Update $\hat{\alpha}^{(i+1)}_{0}$, $\hat{\alpha}^{(i+1)}_{1}$ and $\hat{\alpha}^{(i+1)}_{2}$ using equations \eqref{alph000}, \eqref{alph111} and \eqref{alph222}, respectively.
	\end{algorithmic}
	\text{The process should be continued until the convergence criterion is met. Under the modified pseudo-} \text{log-likelihood function based $a^*$, our process is stopped when $\frac{\abs{Q\(\hat{\alpha}^{(i+1)}_{0}, \hat{\alpha}^{(i+1)}_{1}, \hat{\alpha}^{(i+1)}_{2}\) - Q\(\hat{\alpha}^{(i)}_{0}, \hat{\alpha}^{(i)}_{1}, \hat{\alpha}^{(i)}_{2}\)}}{Q\(\hat{\alpha}^{(i)}_{0}, \hat{\alpha}^{(i)}_{1}, \hat{\alpha}^{(i)}_{2}\)}$} \text{ $ < \epsilon$, where $\epsilon = 10^{-5}$.}
\end{algorithm}

%
%
\subsection{EM algorithm for seven-parameter BB-BVPA distribution}
\label{sec:Pro}

Here we consider the estimation via EM algorithm of the BB-BVPA distribution in presence of location and scale parameters. Let us divided our dataset $\mathcal{D}_2=\{(y_{11},y_{21}),(y_{12},y_{22})\cdots,$ $(y_{1n},y_{2n})\}$ into two part as follows:
\begin{center}
	$ D'_1 =\big \{i : \frac{y_{1i} - \mu_{1}}{\sigma_{1}} < \frac{y_{2i} - \mu_{2}}{\sigma_{2}} \big\}$, $ D'_2 = \big\{ i : \frac{y_{1i} - \mu_{1}}{\sigma_{1}} > \frac{y_{2i} - \mu_{2}}{\sigma_{2}} \big\} $,
\end{center}
and consider
\begin{center}
	$|D'_1| = n_1$, 	$|D'_2| = n_2$, here  $n=n_1+n_2$.
\end{center}
Therefore, the usual log-likelihood function of this BB-BVPA distribution based on dataset $\mathcal{D}_2$ can be written as 
\begin{equation}
	\label{4e1}
	\begin{split}
		L(\mu_{1}, \mu_{2}, \sigma_{1}, \sigma_{2}, \alpha_{0}, \alpha_{1}, \alpha_{2}) =  &~ n\ln(\alpha_{0} + \alpha_{1} + \alpha_{2}) - n\ln(\alpha_{1} + \alpha_{2}) + n_{1}\ln\alpha_{1} + n_{1}\ln(\alpha_{0} + \alpha_{2})\\- &~n_{1}\ln\sigma_{1} - n_{1}\ln\sigma_{2} - (\alpha_{0} + \alpha_{2} + 1)\sum_{i \in D'_{1}} \ln(1 + \frac{y_{2i} - \mu_{2}}{\sigma_{2}})\\ -&~  (\alpha_{1} + 1)\sum_{i \in D'_{1}}^{} \ln(1 + \frac{y_{1i} - \mu_{1}}{\sigma_{1}}) +  n_{2}\ln\alpha_{2} + n_{2}\ln(\alpha_{0} + \alpha_{1})\\ -&~ n_{2}\ln\sigma_{1} - n_{2}\ln\sigma_{2} - (\alpha_{0} + \alpha_{1} + 1)\sum_{i \in D'_{2}}^{} \ln (1 + \frac{y_{1i} - \mu_{1}}{\sigma_{1}})\\ -&~ (\alpha_{2} + 1)\sum_{i \in D'_{2}}^{} \ln(1 + \frac{y_{2i} - \mu_{2}}{\sigma_{2}}). 
	\end{split}
\end{equation} 
We first estimate the location parameters from the marginal distributions of the seven-parameter BB-BVPA distribution and keep them fixed. The minimum of the marginal of the data is used as the estimator of the location parameter. In our EM algorithm, the scale parameters $\sigma_1$, $\sigma_2$ and the shape parameters $\alpha_0$, $\alpha_1$, $\alpha_2$ are updated at every iteration. Since the bivariate log-likelihood function \eqref{4e1} is a discontinuous function with respect to location and scale parameters, gradient descent with respect to bivariate likelihood will not work. Therefore, we estimate scale parameters using the distribution of marginals. At every iteration, we use one-step-ahead gradient descent to estimate $\sigma_1$, $\sigma_2$ based on the likelihood of marginals combined with usual EM steps for other parameters. 

Once we get the estimate of the location and scale parameters, we fix $D'_{1}$ and $D'_{2}$ and then estimate $\alpha_0$, $\alpha_1$, $\alpha_2$ using EM algorithm for the next iteration, which is discussed above. Let us define $z_{1i} = \frac{(y_{1i} - \hat{\mu}_{1})}{\hat{\sigma}_{1}}$ and $z_{2i} = \frac{(y_{2i} - \hat{\mu}_{2})}{\hat{\sigma}_{2}}$, where for $j=1,2$, $\hat{\mu}_{j}$ and $\hat{\sigma}_{j}$ denote estimator of location and scale parameter, respectively. Suppose the estimates of location and scale parameters are exactly the same as the actual location and scale parameters. In that case, this transformed data $(z_{1i}, z_{2i})$ for $i=1,\cdots, n$ are the observations from the three-parameter BB-BVPA distribution. It should be mentioned that the normalized data with respect to the estimated location and scale parameters, the transformation is not going to provide the distribution of normalized data exactly as the three-parameter BB-BVPA distribution. This transformation rather forms some distribution close to $BBBVPA(\alpha_0, \alpha_1, \alpha_2)$. It isn't easy to know the exact distribution. However, we assumed that this transformed data are the observations of $BBBVPA(\alpha_0, \alpha_1, \alpha_2)$.   
Therefore, the pseudo-maximum likelihood estimates of $\alpha_{0}$, $\alpha_{1}$ and $\alpha_{2}$ are same as before, i.e.,
%
\begin{equation} 
	\hat{\alpha}_{0} = \frac{\tilde{n}^{*}_{0} + u_{1} n_{1} + w_{1} n_{2}}{\tilde{n}_{0} a^{*}_{0}  + \sum_{i \in D'_{2}}^{} \ln(1 + z_{1i}) + \sum_{i \in D'_{1}}^{} \ln (1 + z_{2i})}, \label{Mstepabs1}
\end{equation}
\begin{equation} 
	\hat{\alpha}_{1} = \frac{n_{1} + w_{2}n_{2}}{\tilde{n}_{0} a^{*}_{0}  + \sum_{i \in D'_{1} \cup D'_{2}}^{} \ln(1 + z_{1i})}, \label{Mstepabs2}
\end{equation}
and
\begin{equation} 
	\hat{\alpha}_{2} = \frac{n_{2} + u_{2} n_{1}}{\tilde{n}_{0} a^{*}_{0} +  \sum_{i \in D'_{1} \cup D'_{2}}^{} \ln(1 + z_{2i})}. \label{Mstepabs3}
\end{equation}
The key intuition of this algorithm is very similar to stochastic gradient descent. As the number of iterations increases, this EM algorithm tries to force the three parameters $\alpha_0, \alpha_1, \alpha_2$ to pick up the right direction starting from any value, and the one-step-ahead gradient descent algorithm for $\sigma_1$ and $\sigma_2$ gradually ensure them to roam around the actual values. One of the drawbacks of this approach is that it considers too many approximations. However, this algorithm works even for moderately large sample sizes. Sometimes it takes a lot of time to converge or roam around the actual value for some samples. The probability of such events is very low. In such situations, we stop the calculation after $2000$ iterations.
%
Now, the algorithmic steps of our proposed algorithm for seven-parameter BB-BVPA distribution can be written as 
\begin{algorithm}[H]
	\caption{EM algorithm of seven-parameter BB-BVPA distribution.	\label{algo4_ch3}}
    \text{Take minimum of the marginals as estimates of the location parameters, i.e., $\hat{\mu}_1 = \min_i\{y_{1i}\}$ and}
    \text{$\hat{\mu}_2 = \min_i\{y_{2i}\}$. Start with some initial choice of the rest of parameters $\sigma_1$, $\sigma_2$, $\alpha_0$, $\alpha_1$ and $\alpha_2$}.
	\begin{algorithmic}[1]
		\STATE Compute one-step-ahead gradient descent update of scale parameters based on log-likelihood function of the marginals.   
		\STATE Fix $D'_{1}$ and $D'_{2}$ based on the estimated location and scale parameters.
		\STATE Compute $u^{(i)}_{1}$, $u^{(i)}_{2}$, $w^{(i)}_{1}$, $w^{(i)}_{2}$, $\tilde{n}_{0}$ and $a^*$ using $\hat{\alpha}^{(i)}_{0}$, $\hat{\alpha}^{(i)}_{1}$ and $\hat{\alpha}^{(i)}_{2}$.
		\STATE  Update $\hat{\alpha}^{(i+1)}_{0}$, $\hat{\alpha}^{(i+1)}_{1}$ and $\hat{\alpha}^{(i+1)}_{2}$ using equations \eqref{Mstepabs1}, \eqref{Mstepabs2} and \eqref{Mstepabs3}, respectively.
	\end{algorithmic}
	\text{The process should be continued until the convergence criterion is met. Under the modified pseudo-} \text{log-likelihood function based $a^*$, our process is stopped when $\frac{\abs{Q\(\hat{\alpha}^{(i+1)}_{0}, \hat{\alpha}^{(i+1)}_{1}, \hat{\alpha}^{(i+1)}_{2}\) - Q\(\hat{\alpha}^{(i)}_{0}, \hat{\alpha}^{(i)}_{1}, \hat{\alpha}^{(i)}_{2}\)}}{Q\(\hat{\alpha}^{(i)}_{0}, \hat{\alpha}^{(i)}_{1}, \hat{\alpha}^{(i)}_{2}\)}$} \text{ $ < \epsilon$, where $\epsilon = 10^{-5}$.}
\end{algorithm}


\section{Confidence interval}
	\label{s5}
%
This section addresses the problem of computing the confidence interval of unknown model parameters. We obtain confidence intervals (CI) for parameters $\sigma_1$, $\sigma_2$, $\alpha_0$, $\alpha_1$ and $\alpha_2$ by using parametric bootstrap technique \cite[see, e.g.,][]{efron1994introduction} based on $1000$ bootstrap replications. Note that the location parameters $\mu_1$ and $\mu_2$ behave like a threshold. So parametric bootstrap confidence interval for $\mu_1$ and $\mu_2$ does not exist. Therefore, we can use the asymptotic confidence interval for $\mu_1$ and $\mu_2$ separately using the distribution of the estimators. 

Confidence intervals for location parameters can be obtained using the distribution of the estimator of location parameters, $\hat{\mu}_j = \min_i\{y_{ji}\}$, for $j = 1, 2$.  Now the survival function of  $ Z_j = (\hat{\mu}_j - \mu_j)/\sigma_j$ is 
\begin{eqnarray}
	S_{Z_j}(z_j) = \Big\{\frac{\alpha_{0}+\alpha_{1}+\alpha_{2}}{\alpha_{1}+\alpha_{2}}(1+z_j)^{-\alpha_{0}-\alpha_{j}}-\frac{\alpha_{0}}{\alpha_{1}+\alpha_{2}}(1+z_j)^{-\alpha_{0}-\alpha_{1} - \alpha_{2}}\Big\}^n.
\end{eqnarray}
Then 95\% approximate confidence intervals can be written as
\begin{align}
	\label{cie4.}
	X_{(j)} - b_j\hat{\sigma}_j \leq \mu_j \leq X_{(j)} - a_j\hat{\sigma}_j.
\end{align}
We calculate $a_j$ and $b_j$ using the survival functions.  Here we use the the estimates $\hat{\sigma}_1$, $\hat{\sigma}_2$, $\hat{\alpha}_0$, $\hat{\alpha}_1$ and $\hat{\alpha}_2$ to form the confidence intervals. 
%
%
%
However, the above suggested procedure is just an approximate asymptotic confidence interval.  More research is needed to explore better confidence interval than the above suggested one.

\section{Results from the analysis of simulated datasets}
\label{s6}
%
This section presents a simulation study to verify how the proposed procedures behave for different sample sizes of both BB-BVPA models. The numerical results are obtained by using freely available \b{\textsf{R}} software environment \cite{rsoftware}. The codes are run at IIT Guwahati computers with model: Intel(R) Core(TM) i5-6200U CPU 2.30GHz. The codes will be available on request to authors. The average estimates (AE), mean squared error (MSE) and $95\%$ confidence intervals are obtained for both BB-BVPA models. The AEs, MSEs and CIs are obtained for different sample sizes using the EM algorithm based on $1000$ replications. It should be mentioned that all numerical results (in both Sections \ref{s6} and \ref{s7}) for the parameters of both BB-BVPA models are based on a particular initial choice of model parameters, which are reported in both cases, respectively. However, We have tried other initial guesses also, but the average estimates and the corresponding MSEs are the same.

\paragraph*{Three-parameter BB-BVPA distribution} We consider one set of parameter values of $(\alpha_0, \alpha_1, \alpha_2)$: $(2, 0.4, 0.5)$ and $(0.6, 2.2, 2.4)$, and use the following initial choice of model parameters : $\alpha_0 = 1$, $\alpha_1 = 0.2$,  $\alpha_2 = 0.2$. The results are reported in Table~\ref{Table-NR1} and \ref{Table-NR1.}. Estimates are calculated based on the different sample sizes $n = 50, 150, 250, 350, 450$. The results shown here are based on Algorithm~\ref{algo3_ch3}, which works for any choice of the parameters within its usual range. 

\paragraph*{Seven-parameter BB-BVPA distribution}
Here we also consider one set of parameter values of $(\mu_1, \mu_2, \sigma_1, \sigma_2, \alpha_0, \alpha_1, \alpha_2)$: $(0.1, 0.1, 0.8, 0.8, 2, 0.4, 0.5)$  and $(1.0, 2.0, 0.5, 0.5, 0.6, 2.2, 2.4)$. We report AEs and MSEs in Table~\ref{Table-NR2} and \ref{Table-NR2.}.  In this case, we take sample size $n = 450, 550, 1000, 1500$. However, the algorithm works even for smaller sample sizes, although mean square errors are a little higher.  Since the EM algorithm starts after plug-in the estimates of location parameters as minimum of the marginals, we take the initial values of other parameters as $\sigma_1 = 0.4$ $\sigma_2 = 0.4$, $\alpha_{0} = 1$, $\alpha_{1} = 0.2$, and $\alpha_{2} = 0.2$.  It is observed that the average estimates are closer to the actual values of the parameters.  However, the parametric bootstrap confidence interval does not work for location parameters as it never contains the true parameters.  Table~\ref{cit1_ch3} and \ref{cit1_ch3.} represent the confidence intervals based on the procedures described in Section~\ref{s5} for the sample size $450$ and $1000$. The Algorithm~\ref{algo4_ch3} also works for any other choice of parameters within its usual range.
\begin{table}[H]
	\begin{tiny}
		\begin{center}
			\begin{adjustbox}{width=0.80\textwidth}
				\begin{tabular}{|c|c|c|c|c|}\hline
					& Parameters &  $\alpha_0$ & $\alpha_1$ & $\alpha_2$ \\
					n &    &  &  & \\ \hline 
					
					50  & AE &   1.9549  & 0.4640 & 0.5699\\ 
					& MSE &   0.3882  & 0.1699 &  0.2292\\ 
					& Parametric bootstrap  & [0.6381, 2.7367] & [0.0003, 1.4601] & [0.0003, 1.6711] \\ 
					& Confidence interval (CI) &  &  & \\\hline
					
					150  & AE  & 1.9802  & 0.4229	 & 0.5252\\ 
					& MSE  & 0.1797  & 0.0700 & 0.0998\\   
					& Parametric bootstrap  & [1.1441, 2.7102] & [0.0029, 1.0054] & [0.0033, 1.2025]\\ 
					& Confidence interval (CI) &  &  & \\\hline
					
					250 & AE &  1.9876 & 0.4141 & 0.5155\\  
					& MSE & 0.1102 & 0.0421 & 0.0620\\   
					& Parametric bootstrap & [1.3788, 2.6498] & [0.0230, 0.8423] & [0.0326, 1.0031]\\ 
					& Confidence interval (CI) &  &  & \\\hline
					
					350 & AE & 1.9981 & 0.4064 & 0.5059\\  
					& MSE & 0.0851 & 0.0312 & 0.0458\\   
					& Parametric bootstrap & [1.4356, 2.5851] & [0.0638, 0.7576] & [0.0770, 0.9228]\\ 
					& Confidence interval (CI) &  &  & \\\hline
					
					450 & AE & 2.0023 & 0.4028 & 0.5028\\ 
					& MSE & 0.0601 & 0.0216 &  0.0329\\    
					& Parametric bootstrap & [1.5085, 2.5053] & [0.1257, 0.7030] & [0.1537, 0.8709]\\ 
					& Confidence interval (CI) &  &  & \\\hline
				\end{tabular}
			\end{adjustbox}  
			\caption{The average estimates (AE), the mean Square Error (MSE) and parametric bootstrap confidence interval (CI) for $\alpha_0 = 2$, $\alpha_1 = 0.4$ and $\alpha_2 = 0.5$. \label{Table-NR1}}  
		\end{center}
	\end{tiny}
\end{table}

\begin{table}[H]
	\begin{tiny}
		\begin{center}
			\begin{adjustbox}{width=0.90\textwidth}
				\begin{tabular}{|c|c|c|c|c|}\hline
					& Parameters &  $\alpha_0$ & $\alpha_1$ & $\alpha_2$ \\
					n &    &  &  & \\ \hline 
					
					50  & AE &   0.8741 &	2.0617 &	2.2774 \\ 
					& MSE & 0.8971 & 0.5175 & 0.5787\\ 
					& Parametric bootstrap  & [1.04 $\times 10^{-06}$, 2.5677] & [0.7146, 3.2527] & [0.7672, 3.7290] \\ 
					& Confidence interval (CI) &  &  & \\\hline
					
					150  & AE  & 0.6773 & 2.1645 & 2.3685 \\ 
					& MSE  & 0.3139 & 0.1892 & 0.2169\\   
					& Parametric bootstrap  & [3.91$\times 10^{-06}$,	1.9542] & [1.3130, 2.9239] & [1.4180, 3.2148]\\ 
					& Confidence interval (CI) &  &  & \\\hline
					
					250 & AE &  0.6387 & 2.1803 & 2.3921\\  
					& MSE & 0.1998 & 0.1286 & 0.1344\\   
					& Parametric bootstrap & [8.52$\times10^{-06}$, 1.5723] & [1.4583, 2.8085] & [1.6495, 3.0483]\\ 
					& Confidence interval (CI) &  &  & \\\hline
					
					350 & AE & 0.6310 & 2.1888 & 2.3941 \\  
					& MSE & 0.1576 & 0.1031 & 0.1057\\   
					& Parametric bootstrap & [3.06$\times 10^{-05}$, 1.4591] & [1.5468, 2.7953] & [1.7297, 2.9884]\\ 
					& Confidence interval (CI) &  &  & \\\hline
					
					450 & AE & 0.6166 & 2.1973 & 2.3998 \\ 
					& MSE & 0.1285 & 0.0823 & 0.0878\\    
					& Parametric bootstrap & [9.39$\times 10^{-05}$, 1.3325] & [1.6533, 2.7769] & [1.7988, 2.9541]\\ 
					& Confidence interval (CI) &  &  & \\\hline
				\end{tabular}
			\end{adjustbox}  
			\caption{The average estimates (AE), the mean Square Error (MSE) and parametric bootstrap confidence interval (CI) for $\alpha_0 = 0.6$, $\alpha_1 = 2.2$ and $\alpha_2 = 2.4$. \label{Table-NR1.}}  
		\end{center}
	\end{tiny}
\end{table}

\begin{table}[H]
	\begin{center}
		\begin{adjustbox}{width=0.50\textwidth}
		\begin{tabular}{|c|c|c|c|c|c|}\hline
			& Parameters & & $\mu_1$ & $\mu_2$ & $\sigma_1$ \\
			n& &    &  &  &  \\ \hline 
			450 &AE &  & 0.1013 & 0.1011 & 0.8377 \\ 
			& MSE & & 3.0698$\times 10^{-6}$ & 2.5216$\times 10^{-6}$ & 0.0242\\    
			450 & & $\sigma_2$ & $\alpha_0$ & $\alpha_1$ & $\alpha_2$ \\
			& AE & 0.8269 & 1.9138 & 0.4931 & 0.6127\\
			& MSE & 0.0346& 0.1005 & 0.0558 & 0.1187\\  \hline  
			
			550 & Parameters &  & $\mu_1$ & $\mu_2$ & $\sigma_1$ \\
			& AE &  & 0.1012 & 0.1009 & 0.8336\\  
			& MSE & & 2.7214$\times 10^{-6}$ & 1.6559$\times 10^{-6}$& 0.0202\\
			
			550 & & $\sigma_2$ & $\alpha_0$ & $\alpha_1$ & $\alpha_2$ \\
			& AE  & 0.8300 & 1.9286 &  0.4756  & 0.6038\\ 
			& MSE & 0.0276 & 0.0740 & 0.0410 & 0.1032\\    \hline
			
			1000 &  & & $\mu_1$ & $\mu_2$ & $\sigma_1$ \\
			& AE & & 0.1006 & 0.1006 & 0.8313\\   
			& MSE & & 7.1534$\times 10^{-7}$ & 6.5660$\times 10^{-7}$ & 0.0131\\    
			
			1000 & & $\sigma_2$ & $\alpha_0$ & $\alpha_1$ & $\alpha_2$ \\
			& AE  & 0.8072 & 1.9461 &  0.4613 & 0.5613 \\
			& MSE & 0.0184 & 0.0452 & 0.0272 & 0.0605\\  \hline
			
			1500 & Parameters & & $\mu_1$ & $\mu_2$ & $\sigma_1$ \\
			& AE & & 0.1004 &0.1004 & 0.8217\\   
			& MSE & & 4.0696$\times 10^{-7}$ & 2.4771$\times 10^{-7}$ & 0.0089\\  
			
			1500 & & $\sigma_2$ & $\alpha_0$ & $\alpha_1$ & $\alpha_2$ \\ 
			& AE & 0.7968 & 1.9744 & 0.4330 & 0.5210\\
			& MSE & 0.0122 & 0.0288 & 0.0162 & 0.0338\\    \hline
		\end{tabular} 
		\end{adjustbox}
		\caption{The average estimates (AE), the mean square error (MSE)  for $\mu_1 = 0.1$, $\mu_2 = 0.1$, $\sigma_1 = 0.8$, $\sigma_2 = 0.8$, $\alpha_0 = 2$, $\alpha_1 = 0.4$ and $\alpha_2 = 0.5$. \label{Table-NR2}}  
	\end{center}
\end{table}

\begin{table}[H]
	\begin{center}
		\begin{adjustbox}{max width=0.50\textwidth}
			\begin{tabular}{|c|c|c|}\hline
				\diagbox{Parameters}{Sample size}
				& 450  & 1000  \\\hline 
				$\mu_1$ & [0.0959,  0.1013] & [0.0982, 0.1004]  \\
				$\mu_2$ & [0.0966, 0.1012] & [0.0992, 0.1012]  \\
				$\sigma_1$ & [ 0.6039, 1.1443] & [0.6399, 1.0632]  \\ 
				$\sigma_2$ & [0.5415, 1.2182] & [0.5962, 1.1278]  \\
				$\alpha_0$ & [1.3315, 2.5556] & [1.5772, 2.3192]  \\
				$\alpha_1$ & [0.1510, 0.9446] & [0.1954, 0.7821] \\
				$\alpha_2$ & [0.1672, 1.4202] & [0.2258, 1.2264] \\ \hline
			\end{tabular} 
		\end{adjustbox}
		\caption{95\% Confidence intervals (CI) for $\mu_1 = 0.1$, $\mu_2 = 0.1$, $\sigma_1 = 0.8$, $\sigma_2 = 0.8$, $\alpha_0 = 2$, $\alpha_1 = 0.4$ and $\alpha_2 = 0.5$ by the suggested procedure.}
		\label{cit1_ch3}  
	\end{center}
\end{table}

\begin{table}[H]
	\begin{center}
		\begin{adjustbox}{width=0.50\textwidth}
			\begin{tabular}{|c|c|c|c|c|c|}\hline
				& Parameters & & $\mu_1$ & $\mu_2$ & $\sigma_1$ \\
				n& &    &  &  &  \\ \hline 
				450 &AE &  & 1.0004 & 2.0004 & 0.5531 \\ 
				& MSE & & 3.92$\times 10^{-7}$ & 3.50$\times 10^{-7}$ & 0.0229\\    
				450 & & $\sigma_2$ & $\alpha_0$ & $\alpha_1$ & $\alpha_2$ \\
				& AE & 0.5079 & 0.5566 & 2.4438 & 2.4616\\
				& MSE & 0.0262& 0.1317 & 0.4667 & 0.6978 \\  \hline  
				
				550 & Parameters &  & $\mu_1$ & $\mu_2$ & $\sigma_1$ \\
				& AE &  & 1.0003 & 2.0003 & 0.5428\\  
				& MSE & & 3.05$\times 10^{-7}$ & 2.48$\times 10^{-7}$& 0.0195\\
				
				550 & & $\sigma_2$ & $\alpha_0$ & $\alpha_1$ & $\alpha_2$ \\
				& AE  & 0.5061 & 0.5598 &  2.4019  & 2.4595\\ 
				& MSE & 0.0209 & 0.1133 & 0.4195 & 0.5541\\    \hline
				
				1000 &  & & $\mu_1$ & $\mu_2$ & $\sigma_1$ \\
				& AE & & 1.0002 & 2.0002 & 0.5229\\   
				& MSE & & 8.79$\times 10^{-8}$ & 7.21$\times 10^{-8}$ & 0.0109\\    
				
				1000 & & $\sigma_2$ & $\alpha_0$ & $\alpha_1$ & $\alpha_2$ \\
				& AE  & 0.5031 & 0.5714 &  2.3109 & 2.4575 \\
				& MSE & 0.0113 & 0.0698 & 0.2367 & 0.2967\\  \hline
				
				1500 & Parameters & & $\mu_1$ & $\mu_2$ & $\sigma_1$ \\
				& AE & & 1.0001 & 2.0001 & 0.5176\\   
				& MSE & & 3.78$\times 10^{-8}$ & 3.16$\times 10^{-8}$ & 0.0076\\  
				
				1500 & & $\sigma_2$ & $\alpha_0$ & $\alpha_1$ & $\alpha_2$ \\ 
				& AE & 0.5019 & 0.5859 & 2.2797 & 2.4410\\
				& MSE & 0.0078 & 0.0483 & 0.1625 & 0.2070\\    \hline
			\end{tabular} 
		\end{adjustbox}
		\caption{The average estimates (AE), the mean square error (MSE)  for $\mu_1 = 1.0$, $\mu_2 = 2.0$, $\sigma_1 = 0.5$, $\sigma_2 = 0.5$, $\alpha_0 = 0.6$, $\alpha_1 = 2.2$ and $\alpha_2 = 2.4$. \label{Table-NR2.}}  
	\end{center}
\end{table}

\begin{table}[H]
	\begin{center}
		\begin{adjustbox}{max width=0.50\textwidth}
			\begin{tabular}{|c|c|c|}\hline
				\diagbox{Parameters}{Sample size}
				& 450  & 1000  \\\hline 
				$\mu_1$ & [0.9992,  1.0012] & [0.9994, 1.0003]  \\
				$\mu_2$ & [1.9985, 2.0008] & [1.9994, 2.0003]  \\
				$\sigma_1$ & [ 0.3495, 0.9007] & [0.3639, 0.7685]  \\ 
				$\sigma_2$ & [0.2707, 0.9052] & [0.3411, 0.7516]  \\
				$\alpha_0$ & [1.18$\times10^{-05}$, 1.2764] & [0.0312, 1.0893]  \\
				$\alpha_1$ & [1.5130, 4.0461] & [1.5431, 3.4346] \\
				$\alpha_2$ & [1.1541, 4.5128] & [1.5027, 3.5977] \\ \hline
			\end{tabular} 
		\end{adjustbox}
		\caption{95\% Confidence intervals (CI) for $\mu_1 = 1.0$, $\mu_2 = 2.0$, $\sigma_1 = 0.5$, $\sigma_2 = 0.5$, $\alpha_0 = 0.6$, $\alpha_1 = 2.2$ and $\alpha_2 = 2.4$ by the suggested procedure.}
		\label{cit1_ch3.}  
	\end{center}
\end{table}

From these simulation experiments, it is clear that the MSEs of EM estimates are decreased with the increase of sample size $n$, indicating the EM estimators' consistency. 
%
\section{Application: Landslides}
\label{s7}
%
For illustrative purposes, we have investigated extreme precipitation patterns in northern Sweden. It shows how the proposed methods can be used in practice. We use daily accumulated precipitation data (in mm) from Abisko Scientific Research Station in northern Sweden for $100$ years, from 1st January 1913 to 31st December 2012. The data set is taken from \url{https://www.polar.se/stoed-till-polarforskning/abisko-naturvetenskapliga-station/}. Rainfall is a recognized trigger of landslides. Rainfall actually increases the groundwater pressure, which, if very high, can trigger a landslide. Short periods with extreme rain intensities or longer periods of up to three days of more moderate but still high rain intensities can increase the groundwater pressure, which may lead to landslides or debris flows. \citet{guzzetti2007rainfall} propose a threshold relation between duration in hours, $D$, and total rainfall in millimeters, $P$ such that the amount of rainfall below these thresholds is unlikely to cause landslides. For highland climates in central and southern Europe, this threshold relation is
\begin{eqnarray}
	\label{thr}
	P = 7.56\times D^{0.52}.
\end{eqnarray}
Thus, a one-day amount below $39.5$ mm $(7.56\times 24^{0.52})$, or a three-day amount below 69.9 mm $(7.56 \times 72^{0.52})$ are all unlikely to cause a landslide or debris flows.
\par
\begin{figure}[H]
	\centering
	\minipage{0.45\textwidth}
	\includegraphics[height=2.5in]{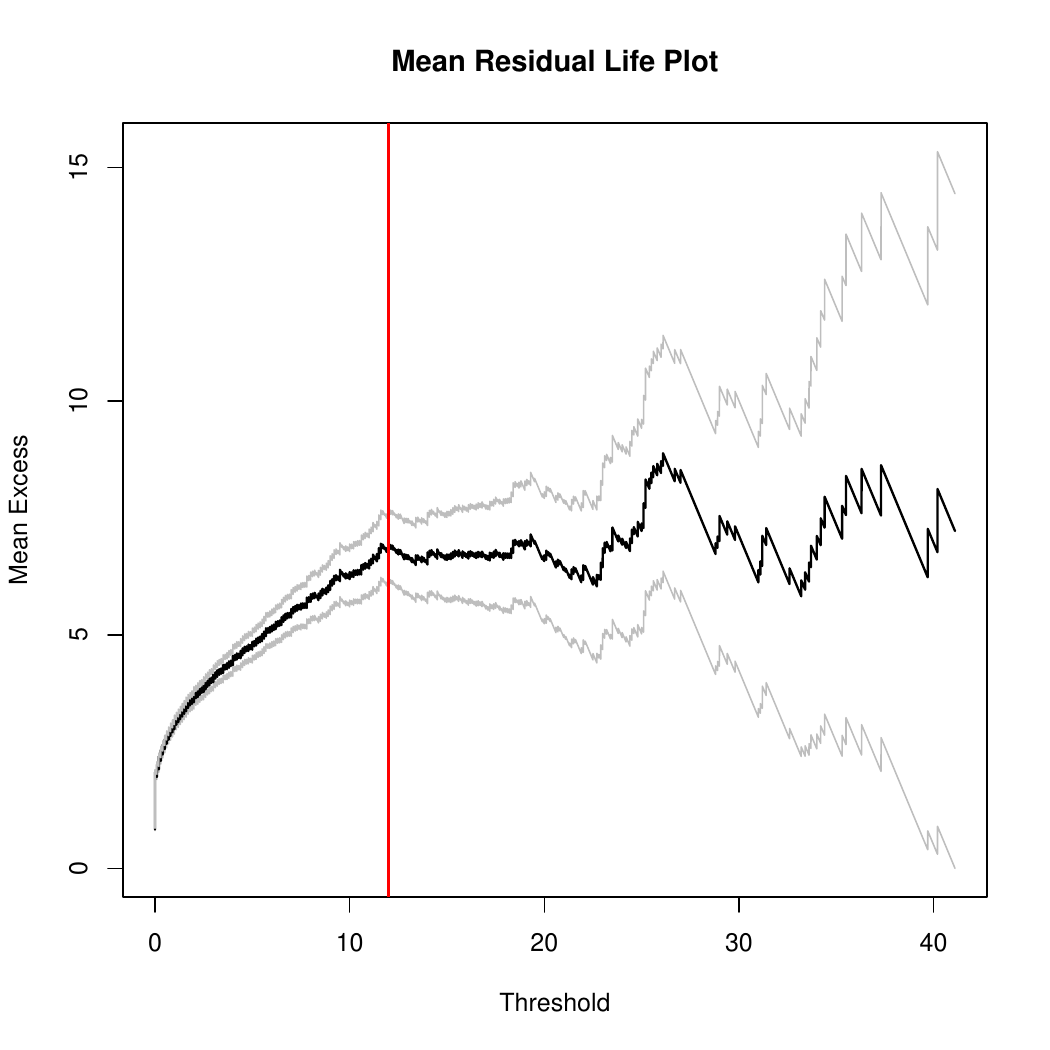}
	\caption{Empirical mean residual life plot of full daily rainfall dataset with threshold $u = 12$ in red.}\label{mrl}
	\endminipage\hfill
	\minipage{0.45\textwidth}
	\includegraphics[height=2.5in]{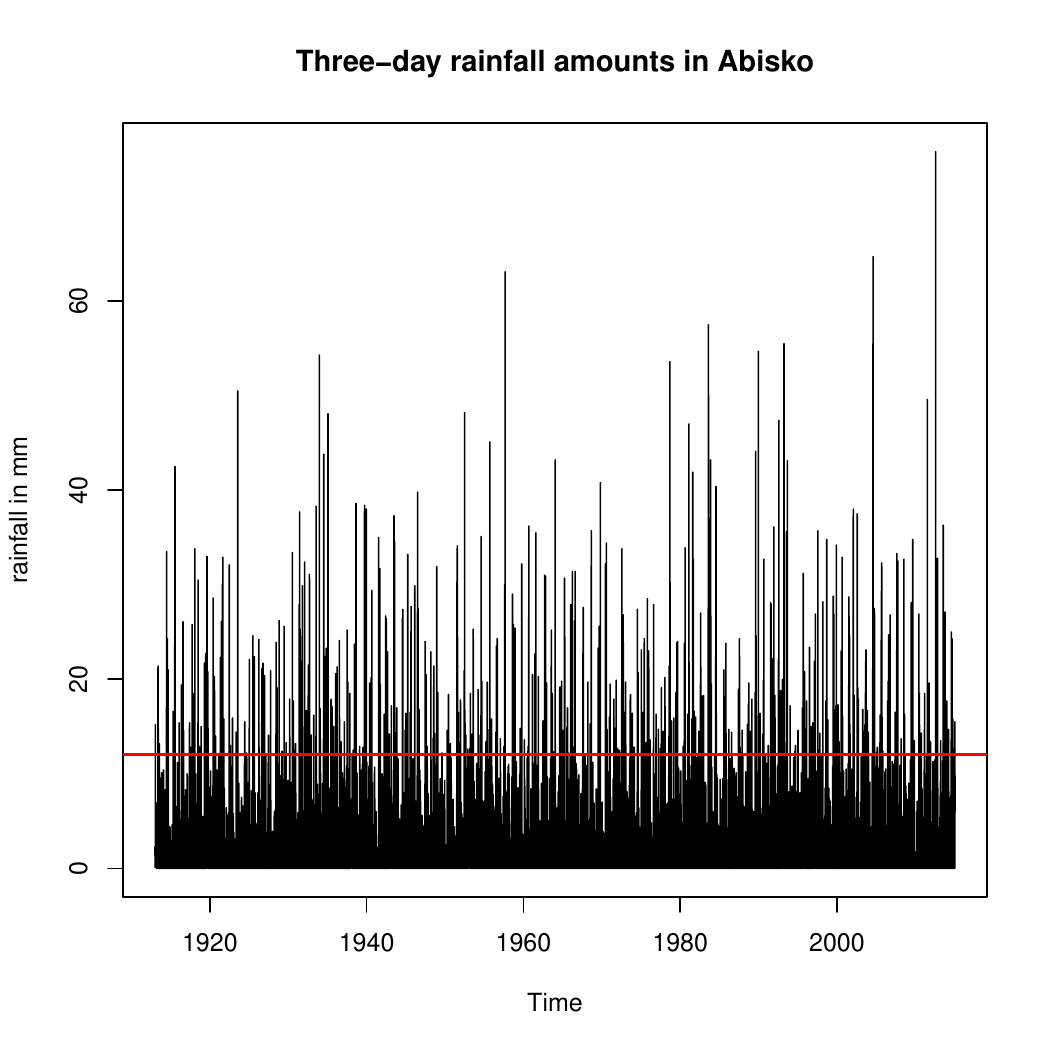}
	\caption{Cumulative three-day precipitation amounts with threshold $u = 12$ in red.}\label{rainfall}
	\endminipage
\end{figure}

 Let us consider \{$P_1$, $P_2$, $\cdots$, $P_N$\} be a long-time series of daily precipitation amounts. Now we construct a dataset $\Y_1$, $\Y_2$, $\cdots$, $\Y_n$ $\in \mathbb{R}^2$, for $n < N$, whose components represent daily and three-day extreme rainfall amounts, respectively, to account for longer periods of moderate rainfall. Based on a mean residual life plot (see Figure~\ref{mrl}), the threshold $u = 12$, which corresponds roughly to the 99\% quantile, is chosen for the daily rainfall amounts \{$P_1$, $P_2$, $\cdots$, $P_N$\}. The cumulative three-day precipitation amounts $P_i + P_{i+1} + P_{i+2}$ for $i \in \{1,2,\cdots, N-2\}$ are shown in Figure~\ref{rainfall}.  The threshold $u$ chosen above is used to extract clusters of data containing extreme episodes; the dataset $\Y_1$, $\Y_2$, $\cdots$, $\Y_n$ are then constructed as follows:
\begin{enumerate}
	\item Find $i$ correspond to the first sum $P_i + P_{i+1} + P_{i+2}$ which exceeds the threshold u chosen above and set $P_{(1)} = \max(P_i, P_{i+1}, P_{i+2})$.
    \item Consider the first cluster $C_{(1)}$ consisting of $P_{(1)}$ plus the five values preceding it and the five values following it.
    \item Let the 1st component of $\Y_1$ be the largest value in $C_{(1)}$,  and the 2nd component the largest sum of three consecutive non-zero values in $C_{(1)}$.
	\item In the same way, find the second cluster $C_{(2)}$ and compute $\Y_2$, starting with the first observation after the cluster  $C_{(1)}$.
\end{enumerate}	
Continuing in the same way, the bivariate dataset $\Y_1$, $\Y_2$, $\cdots$, $\Y_n$ with $n=563$ are obtained.
\par Now we fit our BB-BVPA model under the assumption that all seven parameters are unknown and use our proposed Algorithm~\ref{algo4_ch3} to compute the MLEs of unknown parameters. To start our EM-based algorithm, we need some initial guesses of the unknown parameters $\sigma_{1}$, $\sigma_{2}$, $\alpha_{0}$, $\alpha_{1}$ and $\alpha_{2}$; initial choice are $0.6$, $0.2$, $1.0$, $0.1$ and $1.0$, respectively. Our algorithm provides the estimates of $\mu_1$, $\mu_2$, $\sigma_{1}$, $\sigma_{2}$, $\alpha_{0}$, $\alpha_{1}$ and $\alpha_{2}$ as $5.2$, $12.1$, $6.2723$, $5.5059$, $2.0589$, $0.0025$ and $0.0028$, respectively.  We have also tried some other initial guesses; our proposed algorithm converges to the same point when we use the same stopping criterion.
\par
\begin{figure}[H]
	\centering
	\minipage{0.32\textwidth}
	\includegraphics[height=2.2in]{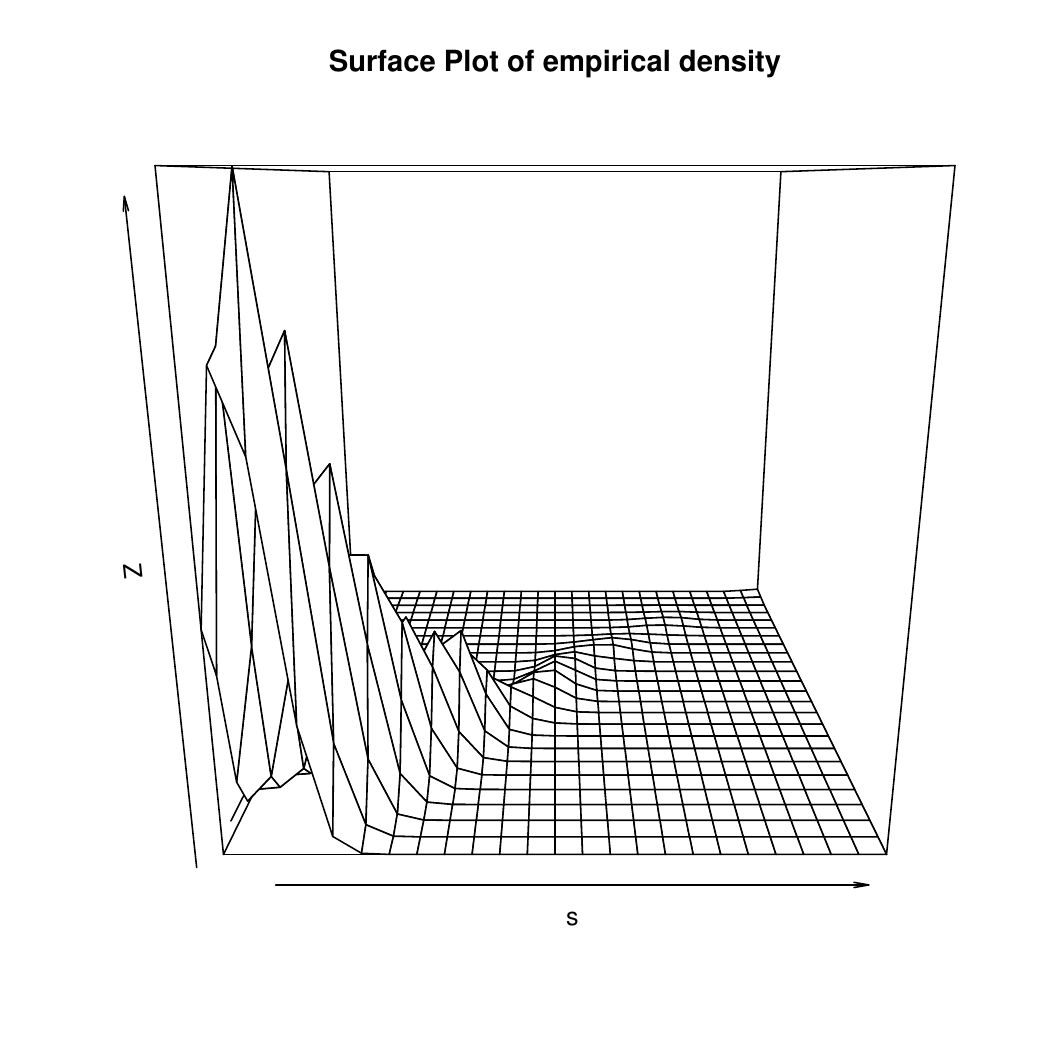}
	\caption{Surface plot of daily and three-day extreme rainfall.}\label{figdensurfaceaxp}
	\endminipage\hfill
	\minipage{0.32\textwidth}
	\includegraphics[height=2.2in]{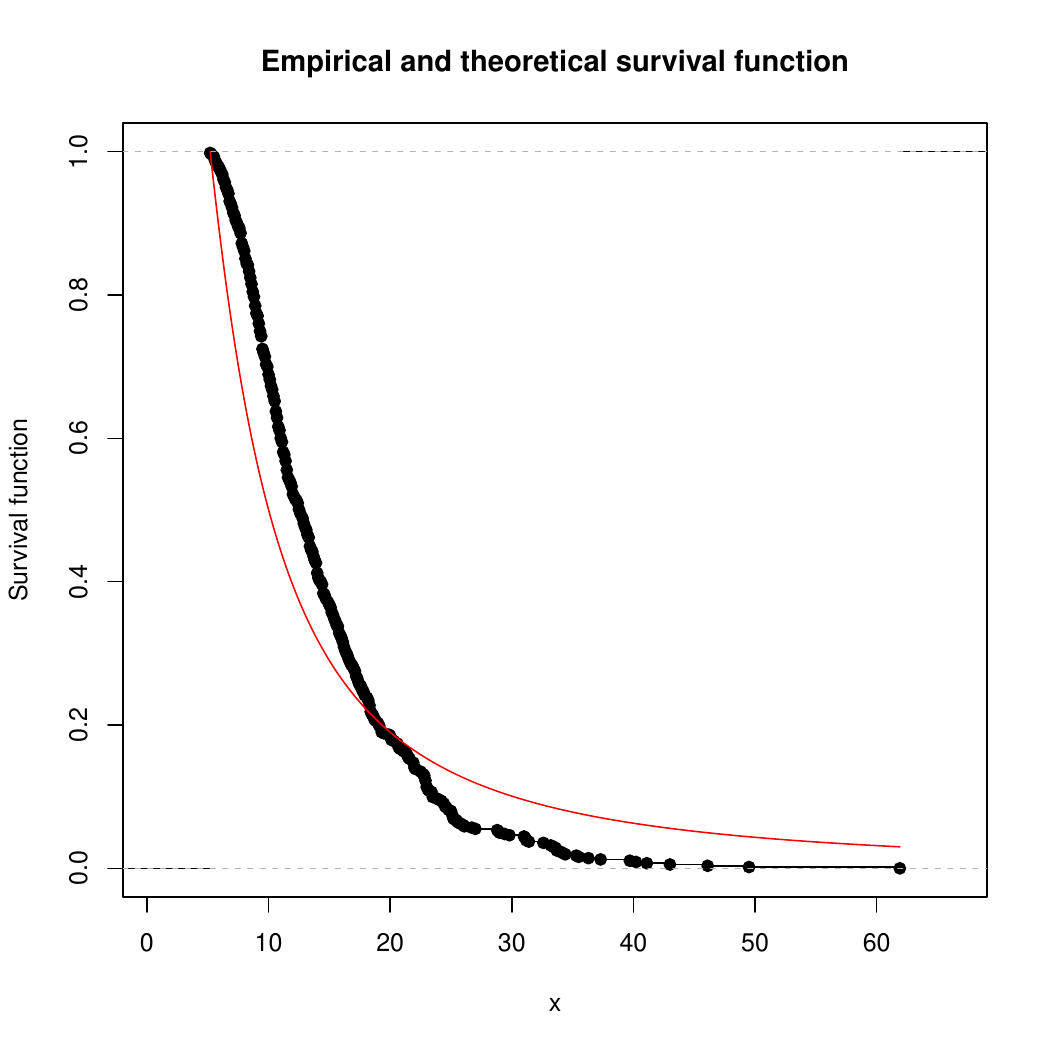}
	\caption{Marginal $Y_1$.}\label{mar1}
	\endminipage\hfill
	\minipage{0.32\textwidth}
	\includegraphics[height=2.2in]{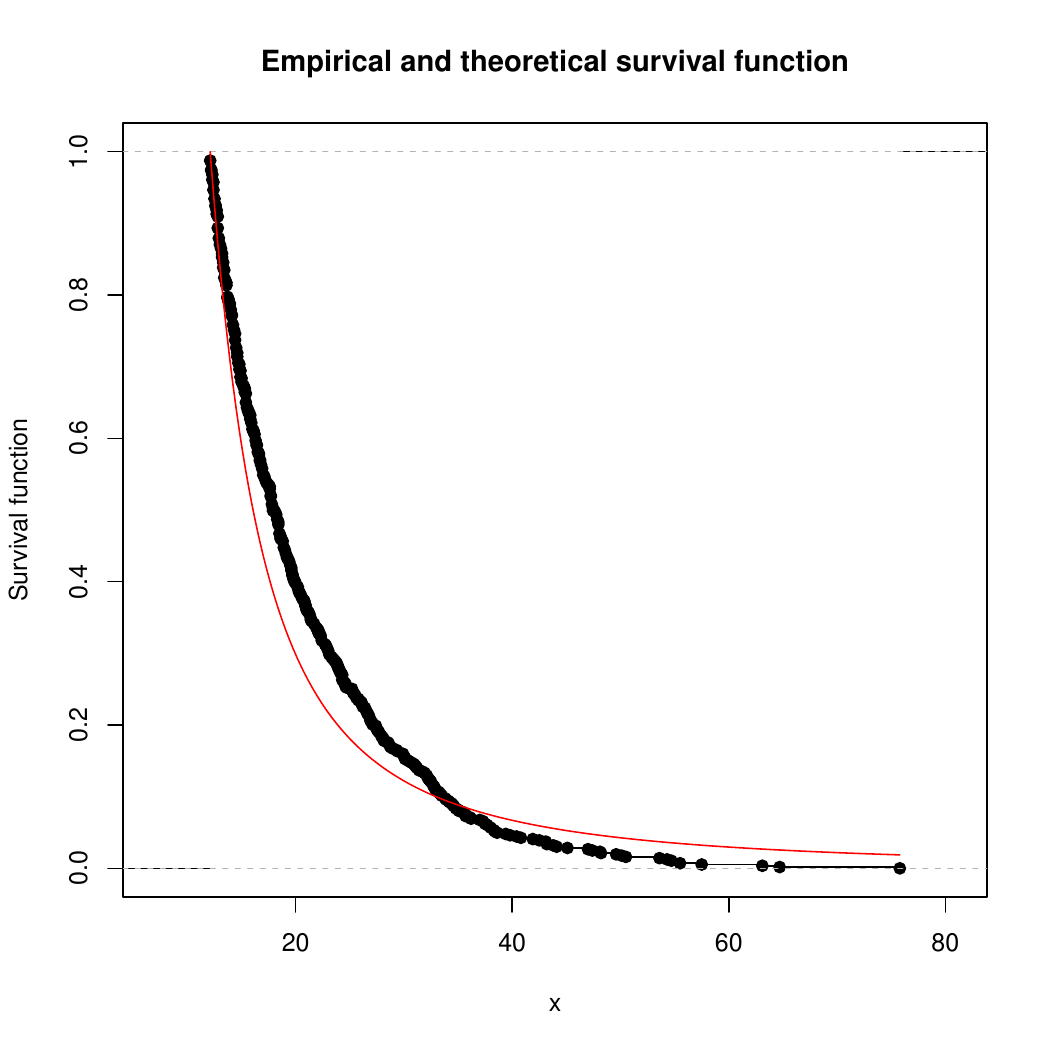}
	\caption{Marginal $Y_2$.}\label{mar2}
	\endminipage
\end{figure}

 A natural question is how well our proposed BB-BVPA model fits the data. Although several goodness of fit tests are available for any arbitrary univariate distribution, not much is available for bivariate distribution. We fit the univariate Pareto type $II$ distribution to the marginals and minimum of standardized variables based on Theorem~\ref{th1} and \ref{th3}. The Kolmogorov-Smirnov (K-S) distances and p-values based on the Chi-square goodness of fit test are reported in Table~\ref{gdtw1_ch3}. It indicates that the BB-BVPA distribution can be used quite effectively for analyzing this bivariate daily and three-day extreme rainfall dataset.
\begin{table}[H]
	\begin{center}
		\begin{adjustbox}{max width=0.50\textwidth}
		\begin{tabular}{|c|c|c|c|}\hline				
			& For $W$  & For marginal $Y_1$ & For marginal $Y_2$ \\\hline 
			K-S distance  & 0.2451 & 0.2273 & 0.20782  \\\hline
			p-value based on & 0.3054 &  0.3335 & 0.3273 \\
			Chi-Square goodness of fit &&&\\\hline
		\end{tabular} 
		\end{adjustbox} 
		\caption{K-S distances and p-value of Chi-square goodness of fit based on daily and three-day extreme rainfall dataset.}
		\label{gdtw1_ch3}  
	\end{center}
\end{table}
Apart from numerical diagnostics, we also verify our assumption by plotting an empirical two-dimensional density plot in Figure~\ref{figdensurfaceaxp}, which resembles closer to the surface of the Block-Basu bivariate Pareto distribution. We fit the empirical survival functions with the marginals of survival of this bivariate Pareto, whose parameters are obtained from the EM algorithm that we have developed. Figure \ref{mar1} and \ref{mar2} show a good fit for both the marginals of BB-BVPA model.
Now our aim is to compare different well-established bivariate Pareto distributions, namely Mardia's type $I$ bivariate Pareto \cite{mardia1962multivariate}, Bivariate Lomax \cite{lindley1986multivariate} and MOBVPA \cite{DeyPaul:2019}. Note that both Mardia's type $I$ and Bivariate Lomax distribution are three-parameter models, whereas both MOBVPA and BB-BVPA are seven-parameter models. From Table~\ref{mdtab}, based on AIC values, it is clear that our proposed BB-BVPA model fits better than all other models.
\begin{table}[H]
	\begin{center}
		\begin{adjustbox}{max width=0.90\textwidth}
			\begin{tabular}{|c|c|c|c|}\hline				
				Bivariate model & Parameters  & Log-likelihood & AIC \\\hline 
				Mardia's type $I$ \cite{mardia1962multivariate}  & $\sigma_1 = 5.2$, $\sigma_2 = 12.1$, $\alpha = 1.4198$ & -3788.638 & 7583.277  \\\hline
				Bivariate Lomax \cite{lindley1986multivariate} & $\alpha_1 = 0.0318$, $\alpha_2 = 0.0224$, $\theta = 2.3303$  & -4486.472 & 8978.943  \\\hline
				MOBVPA \cite{DeyPaul:2019}  & $\mu_1$ = $5.2$, $\mu_2$ = $12.1$, $\sigma_{1}$ = $6.4720$, $\sigma_{2}$ = $5.1660$, & -2382.115 & 4778.230 \\
				& $\alpha_{0}$ = $0.9908$, $\alpha_{1}$ = $0.2533$, $\alpha_{2}$ = $0.2046$ &  & \\\hline	
				BB-BVPA  & $\mu_1$ = $5.2$, $\mu_2$ = $12.1$, $\sigma_{1}$ = $6.2723$, $\sigma_{2}$ = $5.5059$, & -1578.189 & 3170.377 \\
				& $\alpha_{0}$ = $2.0589$, $\alpha_{1}$ = $0.0025$, $\alpha_{2}$ = $0.0028$ &  & \\\hline	
			\end{tabular} 
		\end{adjustbox} 
		\caption{MLEs, the log-likelihood and AIC for four different  bivariate Pareto distributions.}
		\label{mdtab}  
	\end{center}
\end{table}
\par We wish to estimate the probability of a future landslide using formula \eqref{thr}, i.e., we wish to calculate the probability of a landslide occurring as a consequence of one or three days of extreme precipitation for any given year using the parameter estimates given in last row of Table~\ref{mdtab},
\begin{align*}
	& P(Y_1 > 39.5 \cup Y_2 > 69.9) \\ & = 1 - P(Y_1 > 39.5 \cap Y_2 > 69.9)  \\ & = 0.0761.
\end{align*}	
This is higher than the result in \citet{rudvik2012dependence}, who analyzed daily, three-day and five-day extreme precipitation amounts from 1913 to 2008 to estimate the yearly risk. From the dataset, it is clear that the number of extreme rainfall that led the landslide over these 100 (1913-2012) years is seven. This result is very close to the seven or eight extreme rainfalls that we would expect based on our model. A clear limitation of this risk estimate calculation is that the thresholds using the equation \eqref{thr} is not constructed for our particular location. More knowledge about local geological conditions and landslide activity of the specific location may give more precise threshold estimates and hence better risk estimates; it has not been attempted here.
%
\section{Concluding Remarks}
\label{s9}
We have observed the successful implementation of the EM algorithm for the absolute continuous bivariate Pareto distribution. The approximation process runs in multiple stages. Therefore, the algorithm works better for a moderately larger sample size. Even in the case of three parameters, the estimation procedure is not straightforward. This work also shows some innovative approaches to handle the estimation in the case of location and scale parameters. It has an absolutely continuous probability density function, and we have also studied several properties of this distribution. The model was applied to the rainfall data, where the context dictated that extremes of ordered cumulative data were the object of interest. This work can be further used for the discrimination of several models. The Bayesian estimation of this distribution, even without location and scale parameters using Gamma prior and Reference prior, can also be a challenging problem.
%
\section*{Acknowledgements}
\label{s10}
The authors would like to thank the Abisko Scientific Research Station, Abisko, Sweden, for providing the precipitation dataset.
\section*{Declarations}
\label{s11}
\textbf{Data availability} The dataset is publicly available and is taken from \url{https://www.polar.se/stoed-till-polarforskning/abisko-naturvetenskapliga-station/}.
%
\small
\bibliographystyle{plainnat}
\bibliography{7p_bbbvpa_refer}		
\end{document}